\documentclass[preprint,review]{elsarticle}

\usepackage[margin=1.35cm]{geometry}
\usepackage{setspace}

\usepackage{lineno}
\usepackage{xcolor}
\usepackage{url}
\usepackage[hyperindex,breaklinks]{hyperref}
\hypersetup{linkcolor=red, citecolor=red, colorlinks=true}
\usepackage{amsmath}
\usepackage{amsfonts}
\usepackage{amssymb}
\usepackage{amsthm}

\usepackage{graphicx}% Include figure files
\usepackage[capitalise]{cleveref}
\usepackage[caption = false]{subfig}
\usepackage{placeins}

\journal{Physica A}

\usepackage{soul}

\usepackage{graphicx}% Include figure files
\usepackage{dcolumn}% Align table columns on decimal point
\usepackage{bm}% bold math
\usepackage{hyperref}% add hypertext capabilities

\usepackage{url}
\usepackage[hyperindex,breaklinks]{hyperref}
\hypersetup{linkcolor=blue, citecolor=red, colorlinks=true}

\usepackage{xcolor}
\usepackage{xparse}

\begin{document}

\begin{frontmatter}

\title{
Floquet dynamical quantum phase transitions in periodically flux-quenched systems
}

\author{Wen-Hui Nie}

\author{Mei-Yu Zhang}

\author{Lin-Cheng Wang\corref{cor1} }
\ead{wanglc@dlut.edu.cn}
\cortext[cor1]{Corresponding author}

\author{Chong Li}

\address{School of Physics, Dalian University of
	Technology, Dalian 116024, China}
	
\begin{abstract}
Floquet dynamical quantum phase transitions (FDQPTs) reveal many nonequilibrium critical phenomena in periodically driven quantum systems, and their underlying mechanisms have attracted deep attention in recent years. In this paper, we consider an extended XY spin chain under a periodic flux-quench protocol, and demonstrate the effect of the flux difference within each micromotion period on the emergence of FDQPTs, by analyzing physical quantities such as the Loschmidt echo, rate function, and dynamical topological order parameter (DTOP), etc. We also generalize the concept of quench fidelity to periodically driven systems, i.e., Floquet quench fidelity, and discuss the necessary and sufficient conditions for FDQPTs. In contrast to conventional single-quench scenarios, the occurrence of FDQPTs is determined by the requirement of Floquet fidelity condition and segment duration. Our framework may be applied generally to arbitrary periodically driven parameters, providing fundamental insights into how periodic protocols control nonequilibrium phase transitions in quantum many-body systems.
\end{abstract}

                              
\end{frontmatter}

%\linenumbers
\section{Introduction}
Floquet engineering introduces time periodicity to manipulate quantum systems and has emerged as a powerful tool for quantum control\cite{PhysRev.138.B979}. By subjecting quantum matter to periodic driving, one can obtain exotic phases that have no equilibrium counterparts, such as Floquet topological states \cite{PhysRevA.82.033429,kitagawa2010topological,lindner2011floquet} and Floquet time crystals\cite{else2016floquet,zhang2017observation,Sacha_2018,Shukla2025}, etc.
These techniques have been successfully implemented across a wide range of experimental platforms, including ultracold atomic gases\cite{jotzu2014experimental,jimenez2015tunable,wintersperger2020realization}, superconducting qubits\cite{roushan2017chiral}, and photonic systems\cite{wang2019simulating,bao2022light}. The flexible controls provided by synthetic platforms further broaden the scope of research, making it possible to study nonequilibrium quantum dynamical phenomena under Floquet driving.

In various nonequilibrium critical phenomena of quantum many-body systems, the theoretical framework for dynamical quantum phase transitions (DQPTs) has been well established. Originally introduced by Heyl et al.\cite{PhysRevLett.110.135704}, DQPTs are characterized by nonanalytic behavior in the time evolution of quantum systems, manifesting as zeros of the Loschmidt echo in the thermodynamic limit. Subsequently, various diagnostic tools have been developed to detect and characterize DQPTs, including the rate function\cite{heyl2018dynamical,10.1063/1.4969869}, geometric phases\cite{lang2018dynamical}, and dynamical topological order parameters (DTOPs)\cite{budich2016dynamical,bhattacharya2017emergent}. The connection between DQPTs and equilibrium quantum phase transitions has been a long-standing topic of interest, and it has been shown that crossing an equilibrium critical point is neither a necessary nor a sufficient condition for the occurrence of DQPTs\cite{vajna2014disentangling,vajna2015topological,wong2024entanglement,PhysRevB.108.064305,PhysRevA.109.043319,PhysRevA.111.042208,Chenshu,Vijayan2023}. 
In recent years, much research focused on quenched systems controlled by time-independent Hamiltonians, where system parameters are either changed suddenly or varied gradually through ramp protocols\cite{sharma2016slow,zamani2024scaling,baghran2024competition,cao2024exploring,PhysRevB.97.045147,PhysRevB.109.L180303}. However, these insights do not directly address DQPTs in periodically driven systems, while extending the concept of DQPTs to Floquet systems gives rise to a variety of novel physical phenomena\cite{PhysRevB.100.085308,zhou2021floquet,Hamazaki2021,luan2022floquet,PhysRevB.106.094314,PhysRevB.105.094311}.
In periodically driven systems, both the stroboscopic dynamics governed by the effective Floquet Hamiltonian and the micromotion dynamics occurring within a single driving period have been extensively investigated.
The effective Hamiltonian determines the quasi-energy spectrum, while the micromotion contains additional information about the real-time evolution and can exhibit critical phenomena that are not apparent in the stroboscopic picture. 

Implementing periodic driving through piecewise-constant quench protocols offers significant advantages for studying these phenomena. In ultracold atomic gases, lattice parameters can be changed abruptly on timescales much shorter than the intrinsic dynamical timescales, effectively realizing instantaneous quenches\cite{langen2015ultracold,schafer2020tools}. Similarly, in superconducting circuits, magnetic fluxes can be switched rapidly using fast electronic devices\cite{hellings2025calibrating}. 
However, the phase factor is seldom considered in studies of DQPTs\cite{PhysRevB.90.081403,PhysRevB.106.045410,PhysRevB.108.155420,zhang2025dynamics}, and recent studies have demonstrated that flux quenches can induce DQPTs without changing the system’s energy spectrum\cite{nie2024flux,zhang2025phase}.
Motivated by the present research work, in this paper, we will consider an integrable extended XY chain by choosing flux as the parameter for piecewise-constant periodic driving to investigate the micromotion dynamics within a single driving period, and we will discuss the conditions for occurrence of FDQPTs, as well as their differences from single-quench DQPTs,
hoping to provide a useful reference for understanding dynamical critical behavior in periodically driven quantum many-body systems.

The rest of paper is organized as follows. In Sec.~\ref{sec2}, we introduce the extended XY-chain under the protocol of periodic flux quenching, and construct the time evolution operator for the periodically driven system. In Sec.~\ref{sec3}, we investigate FDQPTs in the periodically flux-quenched system by analyzing physical quantities including the Loschmidt echo, rate function, and DTOP to characterize FDQPTs occurring within individual driving periods. 
The Floquet quench fidelity as well as the necessary and sufficient conditions for FDQPTs under both symmetric and asymmetric time allocations have been introduced, and the temporal constraints for the occurrence of FDQPTs under periodic driving are also illustrated from a geometric perspective based on the evolution trajectories on the Bloch sphere.
Finally, we summarize our research work in Sec.~\ref{sec5}.

\section{The extended XY spin chain under Periodic Flux Quench}\label{sec2}  

Consider a periodically driven extended XY spin chain subject to a piecewise constant flux, i.e., the system evolves under a two-segment flux-quench protocol within each driving period $T=T_1+T_2$, where the Hamiltonian is given by 
\begin{eqnarray}
 H( \phi _{\alpha}) =\begin{cases}
		H ( \phi _1  ) ,&		t\in [0, T_1),\\
		H ( \phi _2  ) ,&		t\in [T_1, T),\\
	\end{cases}
\end{eqnarray}
and the driving is rendered periodic through the condition $H(t)=H(t+T)$. Here, the micromotion time $t$ labels the evolution within a single period, and the flux parameters $\phi_\alpha$, $\alpha=1, 2$, denote the quenched value assigned to the corresponding time segments. 
The explicit form of the Hamiltonian $H( \phi _{\alpha}) $ in each segment is ($\hbar=1$)
\begin{eqnarray}
	H(\phi_\alpha) = \sum_{l=1}^L \Big[
	\big(J {-} \gamma \cos \phi_\alpha\big) s_l^x s_{l+1}^x
	+ \big(J {+} \gamma \cos \phi_\alpha\big) s_l^y s_{l+1}^y  {-} \gamma \sin \phi_\alpha \big(s_l^x s_{l+1}^y {+} s_l^y s_{l+1}^x\big)
	{-} \lambda s_l^z \Big],
\end{eqnarray}
where $J$ is the exchange strength, $\lambda$ denotes the strength of the transverse magnetic field, $\gamma$ characterizes the XY anisotropy of the exchange interaction, and $s^j_l=\hbar \sigma ^j_l/2$. Both the anisotropic interaction term and the symmetric off-diagonal exchange term are modulated by the flux parameter 
$\phi_\alpha$. 
Note that our driving protocol differs from the continuous periodic modulations in Refs.\cite{zamani2024scaling,PhysRevB.100.085308,PhysRevB.106.094314}, where the driving contains continuous $\omega t$ time-dependent terms implemented by external microwave pulses, whereas we instead employ segmented flux quenching with piecewise constant $\phi_\alpha$.
As a fine-tuned toy model, it is designed for analytical solvability, yet it captures the essential physics of FDQPTs. From an experimental perspective, the proposed model and its specific quench protocol can be implemented using a negatively charged nitrogen-vacancy (NV) center in a \{100\}-face bulk diamond\cite{PhysRevB.100.085308,PhysRevLett.120.120501}, where the electron spin is manipulated by microwaves generated by an arbitrary waveform generator (AWG), enabling the realization of the piecewise flux quench by varying the phase of the applied microwave pulses in different segments.

The boundary conditions of the system are determined by the eigenvalue parity of the fermion-number operator $N=\sum_{j=1}^L{a_{j}^{\dag}a_j}$. 
Specifically, the fermionic operators satisfy $a_{L+1}\equiv -e^{i\pi N}a_1$ with periodic boundary conditions (PBC) $a_{L+1}=a_1$ for odd $N$, and anti-periodic boundary conditions (APBC) $a_{L+1}=-a_1$ for even $N$\cite{mbeng2024quantum}.
In this work, we always adopt APBC by considering the even fermion number parity, where each segment of $H_{\alpha}$ can be diagonalized. 
By applying the Jordan–Wigner transformation $\sigma^+_j  = \prod_{j' = 1}^{j-1} e^{i \pi n_{j'}} a_j$, $\sigma^-_j = \prod_{j' = 1}^{j-1} e^{i \pi n_{j'}} a^\dagger_j$, $\sigma^z_j = 1 - 2 a^{\dag}_ja_j$, together with the Fourier transforms $a_l = \frac{1}{\sqrt{L}} \sum_k a_k e^{-ik l}$ and $a_l^\dagger = \frac{1}{\sqrt{L}} \sum_k a_k^\dagger e^{ik l}$, the Hamiltonian $H_{\alpha}$  can be represented in momentum space. 
The allowed momenta are restricted to the set $\mathcal{K}^{+} =\{ k| k= (2n-1)\pi/L, n = 1, \cdots, L/2\}$, and in this momentum sector, the Hamiltonian $H (\phi _{\alpha})$ can be expressed as 
\begin{eqnarray}
 H (\phi _{\alpha}) = \sum_{k\in \mathcal{K}^+}
	\Big[ \big(J\cos k{+}\lambda\big)\big( a_{k}^{\dag}a_k {-} a_{-k} a_{-k}^{\dag} \big) 
	{+} i\gamma\sin k \big( e^{-i\phi_\alpha} a_{k}^{\dag} a_{-k}^{\dag}
	{-} e^{i\phi_\alpha} a_{-k} a_k \big) \Big] .
\end{eqnarray}
To extract the nontrivial Bogoliubov block of the Hamiltonian in $k$-mode subspace, we choose the bases 
$\{a_k^\dagger a_{-k}^\dagger |0\rangle,  |0\rangle \}$, under which the Hamiltonian reduces to a nontrivial $2\times2$ block. 
By using a two-component Nambu spinor $\Psi _{k}^{\dag}=( a_{k}^{\dag},a_{-k} )$,   the Hamiltonian $H(\phi_{\alpha})$  can be written in a compact Bogoliubov–de Gennes (BdG) form $H  (\phi _{\alpha}  ) =\sum_{k\in \mathcal{K}^+}{\Psi _{k}^{\dag}}h_{k}(\phi _{\alpha}) \Psi _k$, with the kernel Hamiltonian
\begin{eqnarray}
	{h_{k}(\phi _{\alpha})}{=}
	\left( \begin{array}{cc}
		J\cos k{+}\lambda              &  i\gamma \sin ke^{-i\phi _{\alpha}}   \\
		{-}i\gamma \sin ke^{i\phi _{\alpha}}      &   -J\cos k{-}\lambda
	\end{array}
	\right)  %\nonumber\\
	{=}{\mathbf{d}_{\alpha,k}  \cdot\boldsymbol{\sigma}},
\end{eqnarray}
where $\boldsymbol{\sigma}$
denotes the pseudo-spin Pauli matrices and $\mathbf{d}_{\alpha,k} =\left( \gamma \sin k\sin \phi _{\alpha}, -\gamma \sin k\cos \phi _{\alpha}, J\cos k+\lambda \right)$ acts as an effective magnetic field. 
This BdG representation provides a convenient framework for analyzing the dynamical evolution of the system under periodic flux quenches, where sudden changes of $\phi_\alpha$ 
correspond to abrupt rotations of the effective field $\mathbf{d}_{\alpha,k} $ on the Bloch sphere.
Then, each segment $H(\phi _{\alpha} )$ can be diagonalized via the Bogoliubov transformation
\begin{eqnarray} 
	&&c_k( \phi _{\alpha}) =\cos \frac{\theta _k}{2}a_k-ie^{-i\phi _{\alpha}}\sin \frac{\theta _k}{2}a_{-k}^{\dag}, \nonumber\\
	&&c_{-k}^{\dag}( \phi _{\alpha} ) =-ie^{i\phi _{\alpha}}\sin \frac{\theta _k}{2}a_k+\cos \frac{\theta _k}{2}a_{-k}^{\dag}.
\end{eqnarray}
Here, $\theta _k$ is defined by $\tan \theta _k=\gamma \sin k/ ( J\cos k+\lambda) $. 
This transformation reduces the Hamiltonian to the diagonal form 
\begin{eqnarray}\label{h_phi_alpha}
	H (\phi _{\alpha} )  =\sum_{k\in \mathcal{K}^+}{\xi _k}\left[ 2 c_{k}^{\dag}( \phi _{\alpha} ) c_k( \phi _{\alpha} ) -1 \right],
\end{eqnarray}
where the quasiparticle spectrum is $\xi_k=|\mathbf{d}_{\alpha,k} | =\sqrt{(J\cos k+\lambda)^2+\gamma^2\sin^2k}$. 
A key feature of the flux-quench protocol is that the flux $\phi _{\alpha}$ induces only a unitary rotation of the Bloch vector $\mathbf{d}_{\alpha,k}$ without changing its amplitude. As a result, the spectrum $\xi_k$ remains the same in both segments of the drive. In contrast to conventional quenches, which modify the coupling strength or the magnetic field and therefore change the energy spectrum, our protocol generates micromotion purely through the change in the direction of the Bloch vector $\mathbf{d}_{\alpha,k}$. This geometric character of flux-induced driving is also essential in shaping the resulting DQPTs. 

The Floquet operator $U_k(T)$ in $k$-mode subspace which governs the evolution over one driving period, can be used to define the effective Floquet Hamiltonian $h_{\text{eff},k} $, i.e, 
\begin{eqnarray}\label{U}
	U_k(T) = e^{-ih_{k}(\phi_2) T_2} e^{-ih_{k}(\phi_1) T_1} % \nonumber \\
	=e^{-ih_{\text{eff},k}  T}. 
\end{eqnarray}
The normalized Bloch vector $\mathbf{\hat{n}}_{\text{eff},k}=(n_{\text{eff},x,k}, n_{\text{eff},y,k}, n_{\text{eff},z,k}) $ associated with the effective Hamiltonian 
$h_{\text{eff},k}=\xi_{\text{eff},k}\mathbf{\hat{n}}_{\text{eff},k}\cdot\sigma$, can be obtained as
%\begin{widetext}
	\begin{eqnarray}\label{neff}	
		n_{\text{eff},x,k}&=&\frac{\sin \theta_k}{\sin  ( \xi_{\text{eff},k} T  )}\left[ \sin \phi_1 \sin  ( \xi_k T_1 ) \cos ( \xi_k T_2 ) + \sin \phi_2 \sin  ( \xi_k T_2  ) \cos  ( \xi_k T_1  ) \right. \nonumber \\
		&+& \left. \cos \theta_k ( \cos \phi_1 - \cos \phi_2  ) \sin  ( \xi_k T_1  ) \sin  ( \xi_k T_2  )  \right], \nonumber \\
		n_{\text{eff},y,k} &=&\frac{\sin \theta_k}{\sin  (  \xi_{\text{eff},k} T )} \left[ -\cos \phi_1 \sin  ( \xi_k T_1  ) \cos  ( \xi_k T_2 ) - \cos \phi_2 \sin  ( \xi_k T_2 ) \cos  ( \xi_k T_1  ) \right. \nonumber \\
		&+& \left. \cos \theta_k \left( \sin \phi_1 - \sin \phi_2 \right) \sin \left( \xi_k T_1 \right) \sin \left( \xi_k T_2 \right) \right], \nonumber \\
		n_{\text{eff},z,k} &=&\frac{1}{\sin (  \xi_{\text{eff},k} T  )} \left[ \cos \theta_k \sin  ( \xi_k T  ) + \sin^2 \theta_k \sin  ( \phi_1 - \phi_2 ) \sin ( \xi_k T_1  ) \sin  ( \xi_k T_2  ) \right], % \nonumber \\
	\end{eqnarray}
%\end{widetext}
and the corresponding effective quasienergy spectrum   is given by
\begin{eqnarray}
 \xi_{\text{eff},k}  =%=|\mathbf{n}_{\text{eff},k}|=
	\frac{1}{T}\arccos \big[  \cos  ( \xi_k T_1 ) \cos  ( \xi_k T_2  ) -( \sin ^2\theta _k\cos \left( \phi _1-\phi _2 \right) +\cos ^2\theta _k) \sin  ( \xi_k T_1  ) \sin  ( \xi_k T_2 ) \big].  
\end{eqnarray}
The effective Hamiltonian can also be diagonalized similarly as Eq. (\ref{h_phi_alpha}), 
\begin{eqnarray}\label{diag_Hamiltonian2}
	H_{\text{eff},k} {=}\sum_{k\in \mathcal{K}^+}\xi _{\text{eff},k}  [2 c^{\dag}_{\text{eff},k}  c_{\text{eff},k} {-}1], % \nonumber\\
\end{eqnarray}
with the following Bogoliubov transformation 
\begin{eqnarray} \label{Bogoliubovtrans2}
	&&c_{\text{eff},k}  {=}\cos \frac{\theta _{\text{eff},k}}{2}a_k{-}e^{-i\varphi _{\text{eff},k}}\sin \frac{\theta _{\text{eff},k}}{2}a_{-k}^{\dag}, \nonumber\\
	&&c_{\text{eff},-k}^{\dag}  {=}e^{i\varphi _{\text{eff},k}}\sin \frac{\theta _{\text{eff},k}}{2}a_k{+}\cos \frac{\theta _{\text{eff},k}}{2}a_{-k}^{\dag},
\end{eqnarray}
where $\theta _{\text{eff},k}$ and $\varphi _{\text{eff},k}$ are defined as 
\begin{eqnarray}
	\tan \theta _{\text{eff},k}{=}\frac{\sqrt{n_{\text{eff},x,k}^{2}{+}n_{\text{eff},y,k}^{2}}}{n_{\text{eff},z,k}}, %\nonumber\\
	\tan \varphi _{\text{eff},k}{=}\frac{n_{\text{eff},y,k}}{n_{\text{eff},x,k}}.
\end{eqnarray} 
Furthermore, the quasiparticle operators corresponding to each segment can be expressed in terms of those of the effective Hamiltonian as
%\begin{widetext}
	\begin{eqnarray}
		c_k( \phi _{\alpha} ) &{=}&\Big[ \cos \frac{\theta _k}{2}\cos \frac{\theta _{\text{eff},k}}{2}{+}i\sin \frac{\theta _k}{2}\sin \frac{\theta _{\text{eff},k}}{2}e^{i\left( \varphi _{\text{eff},k}{-}\phi _{\alpha} \right)} \Big] c_{\text{eff},k}  \nonumber \\
		&{+}&\Big[ \cos \frac{\theta _k}{2}\sin \frac{\theta _{\text{eff},k}}{2}e^{{-}i\varphi _{\text{eff},k}}{-}i\sin \frac{\theta _k}{2}\cos \frac{\theta _{\text{eff},k}}{2}e^{{-}i\phi _{\alpha}} \Big] c_{\text{eff},-k}^{\dag}  , \nonumber \\
		c_{-k}^{\dag}( \phi _{\alpha} )&=&
		\Big[ -\cos \frac{\theta _k}{2}\sin \frac{\theta _{\text{eff},k}}{2}e^{i\varphi _{\text{eff},k}}-i\sin \frac{\theta _k}{2}\cos \frac{\theta _{\text{eff},k}}{2}e^{i\phi _{\alpha}} \Big] c_{\text{eff},k} \nonumber \\
		&+&\Big[ \cos \frac{\theta _k}{2}\cos \frac{\theta _{\text{eff},k}}{2}-i\sin \frac{\theta _k}{2}\sin \frac{\theta _{\text{eff},k}}{2}e^{i( \phi _{\alpha}-\varphi _{\text{eff},k} )} \Big] c_{\text{eff},-k}^{\dag}  .
	\end{eqnarray}
%\end{widetext}
This transformation establishes a connection between the segment Hamiltonian quasiparticles and those of the effective Hamiltonian, allowing the system's periodic dynamics under a multi-step quench to be described within a unified quasiparticle framework. For subsequent analysis, $|\psi_k(t)\rangle = U_k(t) |\psi_k(0)\rangle$, and 
the time-evolution operator within a single period is given by
\begin{eqnarray}
	U_k(t) {=} \left\{
	\begin{aligned} %{array}{l}
		&e^{{-}i h_{k}(\phi_1) t},   \ \ \ \ \ \ \ t \in [0, T_1), \\   
		&e^{{-}i h_{k}(\phi_2) (t {-} T_1)} e^{{-}i h_{k}(\phi_1) T_1},  t \in [T_1,T ).   
	\end{aligned}   \right. % \nonumber\\
\end{eqnarray}

%%%%%%%%%%%%%%%%%%%%%%%
\section{Floquet dynamical quantum phase transitions in the periodically flux-quenched system}\label{sec3}

Under periodic driving, when FDQPTs occur, they appear within each driving period and repeat periodically without rapid decay over time. Therefore, we will focus on the micromotion dynamics within one period flux quenching, and shows emergence of FDQPTs during the evolution. In this study, we always assume that the system is initially prepared in the ground state of the effective Hamiltonian $| G_{\text{eff}}  \rangle$, which satisfies $c_{\text{eff},k} | G_{\text{eff}} \rangle =0$.   
Such state of the system always ensures the same dynamical properties of micromotion for each driving period, and the overall evolution remains stable with integer multiples of the driving period.

\subsection{Loschmidt echo}
\begin{figure}
	\centering
	{\includegraphics[width=0.286\columnwidth, height=0.22\columnwidth]{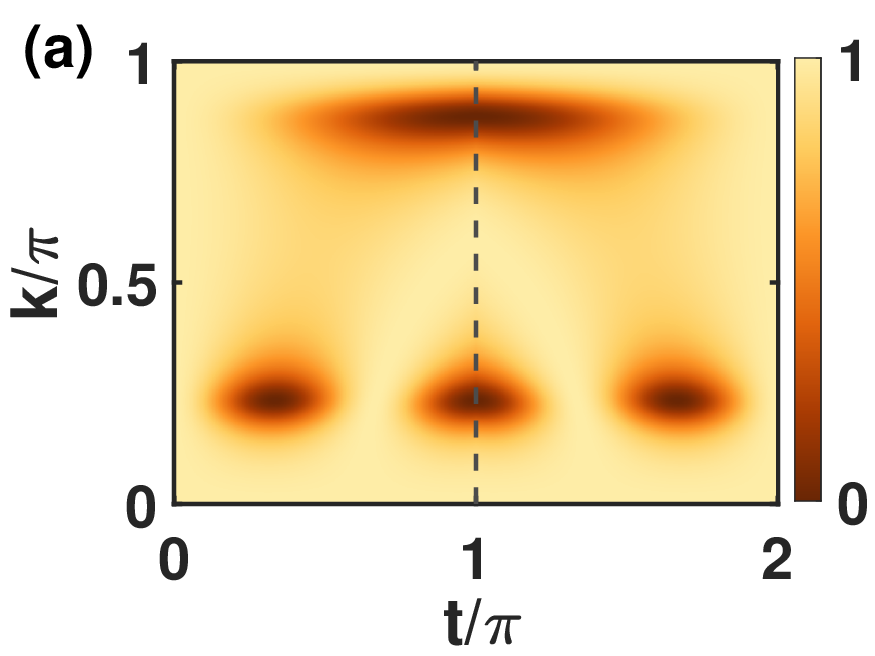}}
	{\includegraphics[width=0.286\columnwidth, height=0.22\columnwidth]{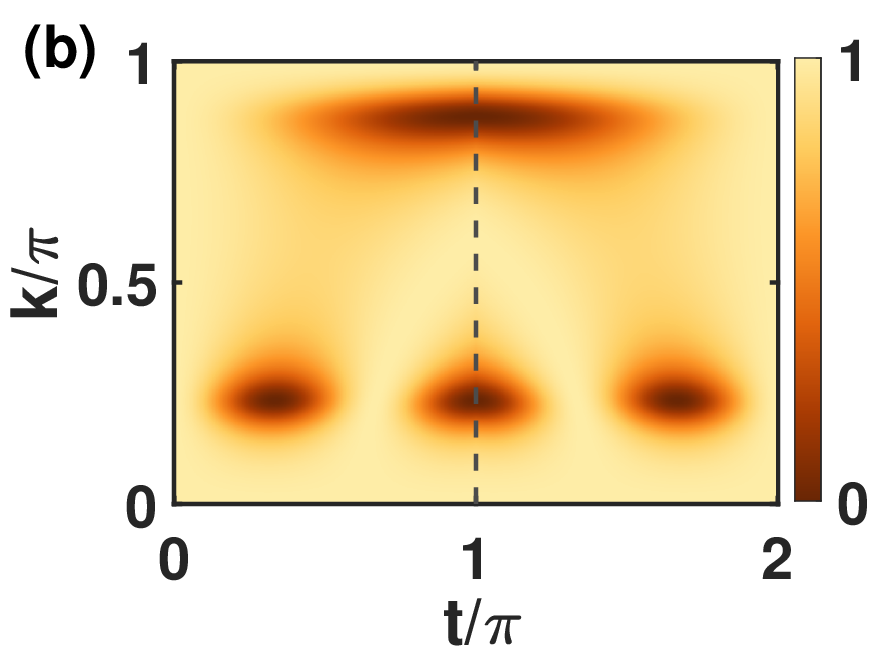}}
	{\includegraphics[width=0.286\columnwidth, height=0.22\columnwidth]{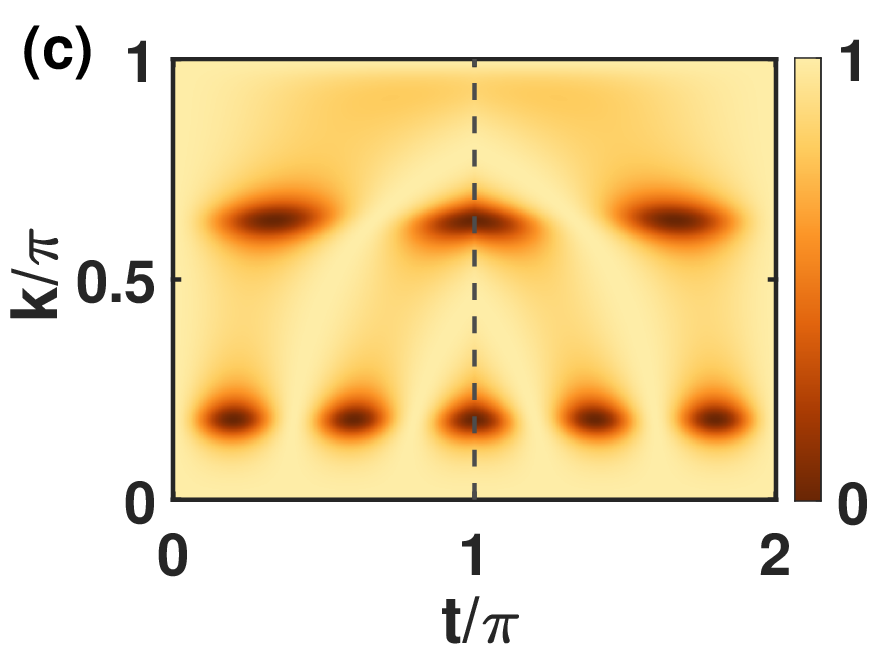}}
	{\includegraphics[width=0.286\columnwidth, height=0.22\columnwidth]{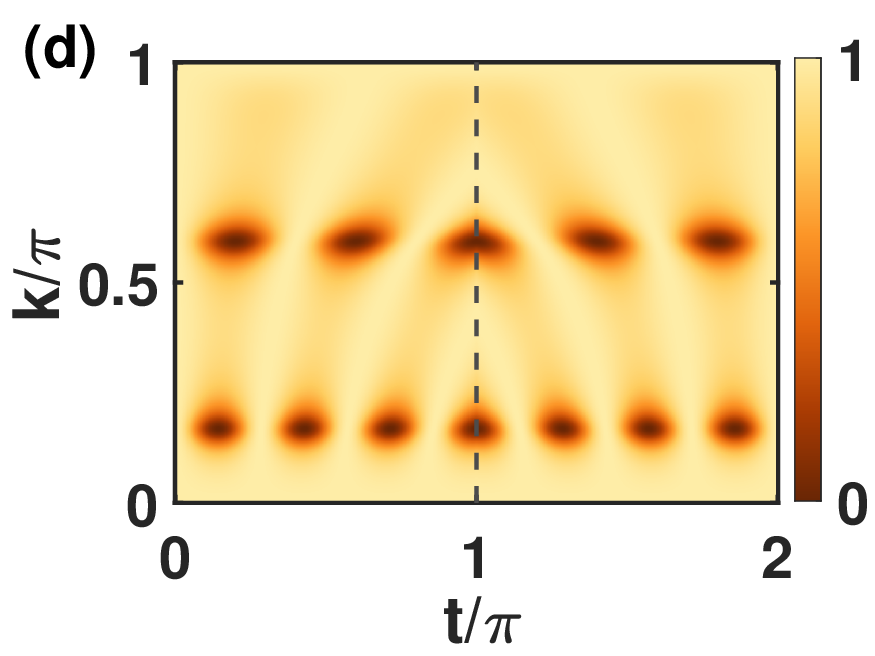}}
	{\includegraphics[width=0.286\columnwidth, height=0.22\columnwidth]{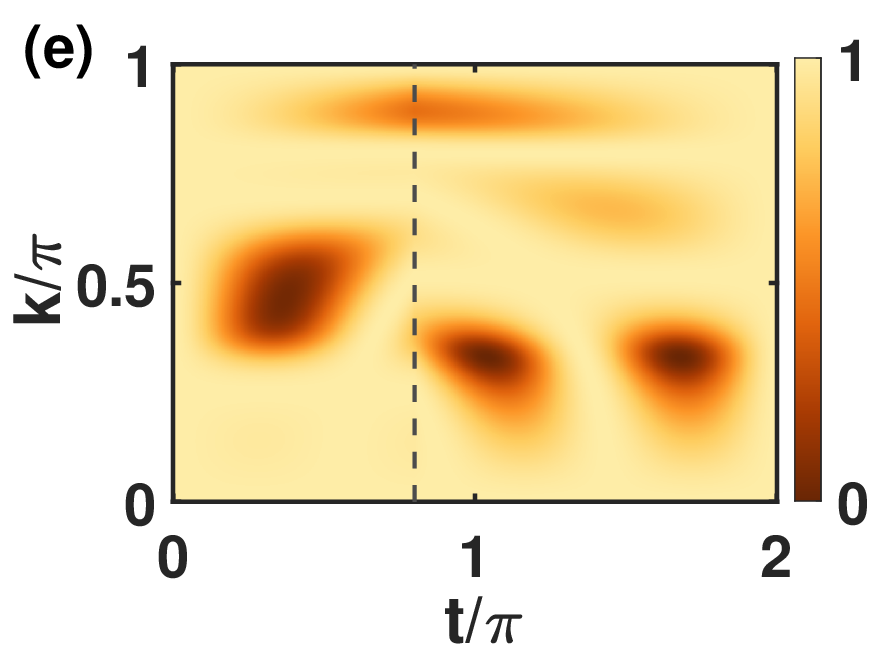}}
	{\includegraphics[width=0.286\columnwidth, height=0.22\columnwidth]{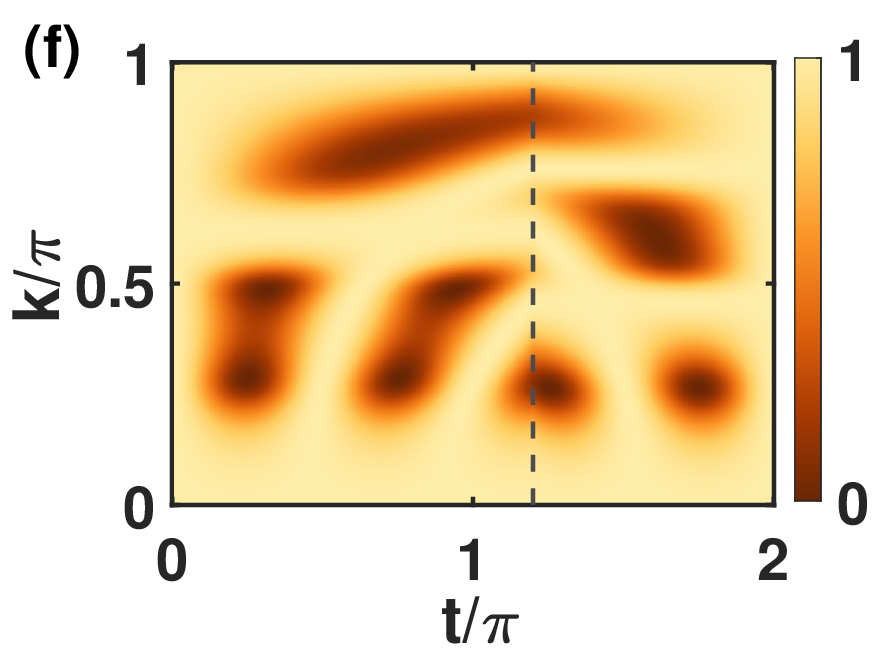}}
	\caption{The single-mode Loschmidt echo $\mathcal{L}_k\left( t \right) $ of the extended XY spin chain under periodically quenched flux, as a function of time $t$ and momentum $k$.  We have set $J = 1$, $\gamma = 1$, and $N = 1000$ in the Numerical calculation.Panels (a)–(d) show the  case of symmetric driving, $T_1 = T_2 = \pi$, while panels (e) and (f) illustrate the asymmetric driving cases with $T_1 = 0.8\pi$, $T_2 = 1.2\pi$, and $T_1 = 1.2\pi$, $T_2 = 0.8\pi$, respectively. In the thermodynamic limit, the zeros of the Loschmidt echo indicate the occurrence of Floquet dynamical quantum phase transitions within each driving period.  
		(a) $\lambda=0.6, \phi_1=0, \phi_2=\pi/4$;  
		(b) $\lambda=0.6, \phi_1=\pi/4, \phi_2=\pi/2$;
		(c) $\lambda=1.6, \phi_1=\pi/2, \phi_2=0$;    
		(d) $\lambda=2.6, \phi_1=3\pi/2, \phi_2=\pi/2$;
		(e) $\lambda=0.8, \phi_1=\pi, \phi_2=0$;    
		(f) $\lambda=1.2, \phi_1=\pi, \phi_2=0$.
	} 
	\label{fig_FLE}
\end{figure}
During one period evolution of the system, the first segment of the evolution under the periodic flux-quench protocol, for $t \in [0, T_1)$, the Loschmidt amplitude is given by
\begin{eqnarray}\label{G1t}
	\mathcal{G}_1( t ) &=&\prod_{k\in \mathcal{K}^+}{\mathcal{G}_{1,k}(t)}=\prod_{k\in \mathcal{K}^+}{\left<  G_{\text{eff}}  |e^{-ih_1t}| G_{\text{eff}} \right>} \nonumber \\
	&=&\prod_{k\in \mathcal{K}^+}{\left[ 1+( e^{-2i\xi _kt}-1) \left| u_{1,k} \right|^2 \right]}e^{i\xi _kt},
\end{eqnarray}
and for the second segment, $t \in [T_1,T )$, the Loschmidt amplitude is given by
\begin{eqnarray}\label{G2t}
	\mathcal{G}_2\left( t \right) &=&\prod_{k\in \mathcal{K}^+}{\mathcal{G}_{2,k}\left(t \right)}=\prod_{k\in \mathcal{K}^+}{\langle  G_{\text{eff}}  |e^{{-}ih_2(t{-}T_1)}e^{{-}ih_1T_1}| G_{\text{eff}}  \rangle}  \notag\\
	&=&\prod_{k\in \mathcal{K}^+}{\left[ 1{+}( e^{-2i\xi _k( t{-}T)}{-}1) \left| u_{2,k} \right|^2 \right]}e^{i\xi _k\left( t-T \right)}e^{i\xi _{\text{eff},k}T}, \nonumber\\
\end{eqnarray}
where the parameter $u_{\alpha,k}$  is defined as % characterizes
\begin{equation}\label{u}
	u_{\alpha,k}{=}\cos \frac{\theta _k}{2}\sin \frac{\theta _{\text{eff},k}}{2}e^{-i\varphi _{\text{eff},k}}{-}i\sin \frac{\theta _k}{2}\cos \frac{\theta _{\text{eff},k}}{2}e^{-i\phi _\alpha}.
\end{equation}
Consequently, the Loschmidt echo $\mathcal{L}\left( t \right) $ can be expressed over the entire period as follows, 
\begin{equation}
\mathcal{L}\left( t \right) =\prod_{k\in \mathcal{K}^+}{\mathcal{L}_k\left( t \right)}=\left\{ \begin{aligned}
	\prod_{k\in \mathcal{K}^+}{\mathcal{L}_{1,k}\left( t \right)}&=\prod_{k\in \mathcal{K}^+}{\left| 1+\left( e^{-2i\xi _kt}-1 \right) \left| u_{1,k} \right|^2 \right|^2},t\in \text{[0,}T_1\text{),}\\
	\prod_{k\in \mathcal{K}^+}{\mathcal{L}_{2,k}\left( t \right)}&=\prod_{k\in \mathcal{K}^+}{\left| 1+\left( e^{-2i\xi _k\left( t-T \right)}-1 \right) \left| u_{2,k} \right|^2 \right|^2},t\in \text{[}T_1,T\text{)}.\\
\end{aligned} \right. 
\end{equation}

Fig.~\ref{fig_FLE} shows the behavior of the single-mode Loschmidt echo $\mathcal{L}_k\left( t \right) $ under the periodic flux-quench protocol in the extended XY spin chain, and FDQPTs emerge depending on the flux difference $\Delta\phi=\phi_1-\phi _2$ between the two quench segments.
Comparing Fig.~\ref{fig_FLE}(a) and (b), we can find that 
when the flux difference $\Delta\phi$ is fixed, the subsequent evolution of the single-mode Loschmidt echo $\mathcal{L}_k\left( t \right) $ follows identical trajectories, indicating that the system’s dynamical response is determined by the relative flux variation rather than by the individual flux values.
The symmetric cases ($T_1 = T_2 = \pi$) are displayed in Fig.~\ref{fig_FLE}(a)–(d), and the asymmetric cases ($T_1 = 0.8\pi, T_2 = 1.2\pi$ and $T_1 = 1.2\pi, T_2 = 0.8\pi$) are shown in Fig.~\ref{fig_FLE}(e)–(f). 
These results demonstrate that both flux modulation and the segment durations determine the occurrence of FDQPTs, reflecting their combined effect on the system dynamics under the driving protocol. 
Fig.~\ref{fig_FLE}(a), (c) and (d) also show that the number of zeros of $\mathcal{L}_k\left( t \right) $ increases with the strength of transverse field $\lambda$, indicating that stronger transverse fields may lead to more frequent emergence of Floquet DQPTs.
The combined effects of flux difference, transverse field strength, and segment duration allocation on the occurrence of Floquet DQPTs are to be further investigated in the subsequent section.

\subsection{Floquet quench fidelity}
The concept of quench fidelity, which has been proposed to establish the relation between the dynamical quantum phase transition and equilibrium quantum phase transition, is defined by the overlap between the ground states of the pre- and post-quench Hamiltonians in the case of a single quench\cite{Chenshu}.  In order to investigate periodically driven systems under Floquet dynamics, we extend the concept of quench fidelity to our periodic quenched system, by defining the Floquet quench fidelity as the overlap between the ground state of the effective Floquet Hamiltonian $H_{\text{eff}}$ and the ground states of each segment of quenched Hamiltonians $H(\phi_{\alpha})$, i.e., 
\begin{eqnarray}
	F_{\alpha}=\left| \langle G_{\text{eff}} | G_{\alpha} \rangle \right|=\prod_{k\in \mathcal{K}^+}{F_{\alpha,k}}. 
\end{eqnarray}
Here, the ground state of the effective Hamiltonians is $\left| G_{\text{eff}} \right> {=}\prod_k{\left( \cos \frac{\theta _{\text{eff},k}}{2}{+}e^{-i\varphi _{\text{eff},k}}\sin \frac{\theta _{\text{eff},k}}{2}a_{k}^{\dag}a_{-k}^{\dag} \right) \left | 0_k,0_{-k} \right>}$ and $|G_{\alpha}\rangle=\prod_{k'}\left(\cos\frac{\theta_{k'}}{2}+ i e^{-i\phi_{\alpha}} \sin\frac{\theta_{k'}}{2}\,a_{k'}^{\dagger} a_{-k'}^{\dagger}\right)\,|0_{k'},0_{-k'}\rangle$ is the ground state of the post-quench Hamiltonian under the segmented flux-quench protocol. 
Then, the $k$ mode of the Floquet quench fidelity can be obtained as
\begin{eqnarray}
	F_{\alpha,k}{=}\left| \cos \frac{\theta _{\text{eff},k}}{2}\cos \frac{\theta _k}{2}{+}i\sin \frac{\theta _{\text{eff},k}}{2}\sin \frac{\theta _k}{2}e^{i\left( \varphi _{\text{eff},k}{-}\phi _{\alpha} \right)} \right|.  \nonumber\\
\end{eqnarray}  
By extending the Loschmidt echo to the complex-time plane, 
the Fisher zeros that cross the imaginary-time axis determine the critical times of the two-stage protocol, 
\begin{equation} \label{Tc}  
	\left\{  
	\begin{aligned}  
		&t_{1,c}=\frac{\left( 2n+1 \right) \pi}{2 \xi_k } , & t \in [0, T_1) \\
		&t_{2,c}=\frac{\left( 2n+1 \right) \pi}{2 \xi_k }+T ,  &t \in [T_1, T) .  
	\end{aligned}  
	\right.   
\end{equation}
and the general expression for the minimum of the single-mode Loschmidt echo $\mathcal{L}_{\alpha,k}^{*}$ during the two-stage quench process is given by
\begin{eqnarray}
	\mathcal{L}_{\alpha,k}^{*}=\left\{ \begin{aligned}%{l}
		&\mathcal{L}_{1,k}(\frac{\pi}{2\xi _k} ),    \\
		&\mathcal{L}_{2,k} (\frac{\pi}{2\xi _k}+T ),\\
	\end{aligned} \right. 
\end{eqnarray}
where the case $n=1$ as specified in Eq. (\ref{Tc}) has been considered. 
In the case of $k=k_c$, $\mathcal{L}_{\alpha,k_c}^{*}$ vanishes, and FDQPTs occur.
Accordingly, the relationship between the minimum of the single-mode Loschmidt echo $\mathcal{L}_{\alpha,k}^{*}$ in each segment and the Floquet quench fidelity $F_{\alpha,k}$ is
\begin{eqnarray}
	\mathcal{L}_{\alpha,k}^{*}=( 1-2F_{\alpha,k}^{2} )^2,
\end{eqnarray}
which relates $\mathcal{L}_{\alpha,k}^{*}$ in each driving segment with the single-mode Floquet quench fidelity $F_{\alpha,k}$. 
Obviously, the zero of $\mathcal{L}_{\alpha,k_c}^{*}$ occurs in the case of $F_{\alpha,k_c} = \sqrt{2}/2$,
which allows us to identify the occurrence of FDQPTs in each $k$-sector by examining the overlap between the ground state of the effective Floquet Hamiltonian and the ground state of the quenching Hamiltonian in each segment.
It highlights the Floquet quench fidelity as an essential physical quantity that bridges equilibrium and dynamical critical behavior. 
\begin{figure}
	\centering
	{\includegraphics[width=0.24\columnwidth, height=0.18\columnwidth]{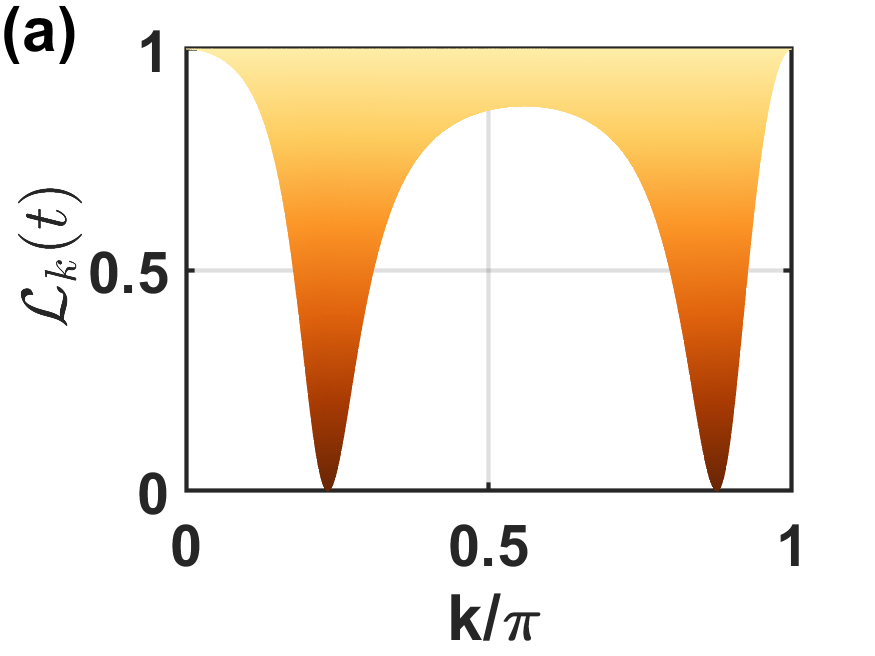}}
	{\includegraphics[width=0.24\columnwidth, height=0.18\columnwidth]{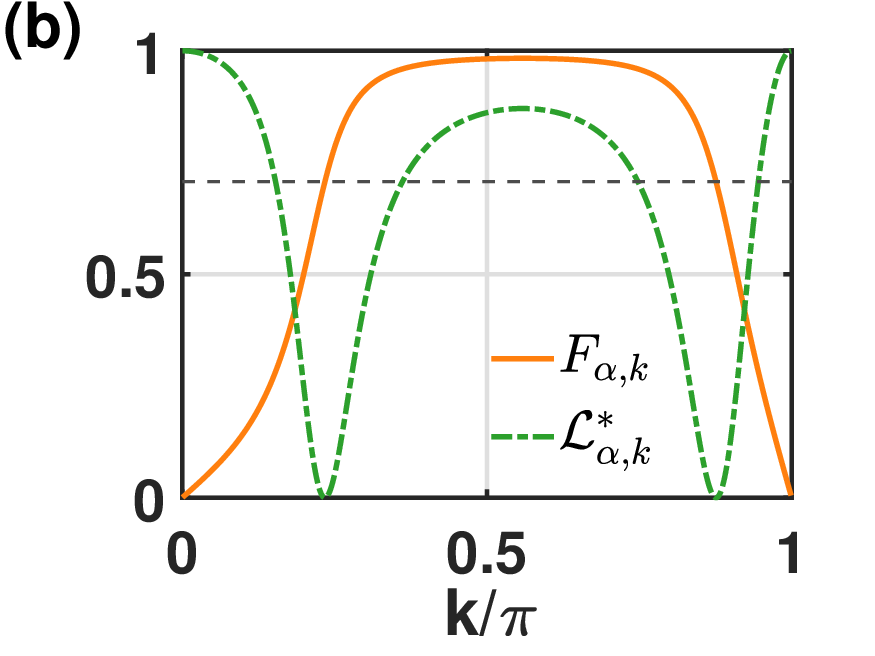}}
	{\includegraphics[width=0.24\columnwidth, height=0.18\columnwidth]{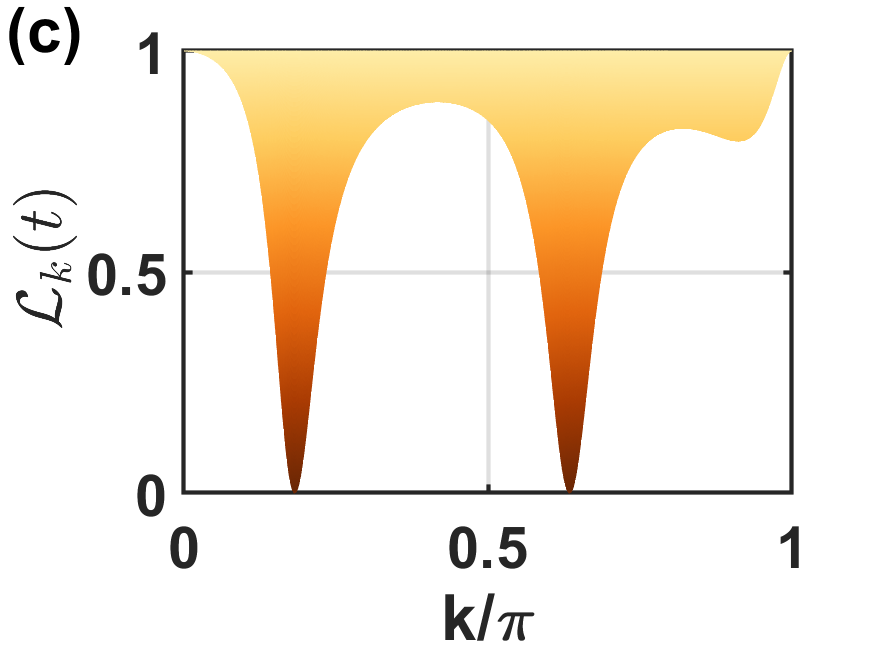}}
	{\includegraphics[width=0.24\columnwidth, height=0.18\columnwidth]{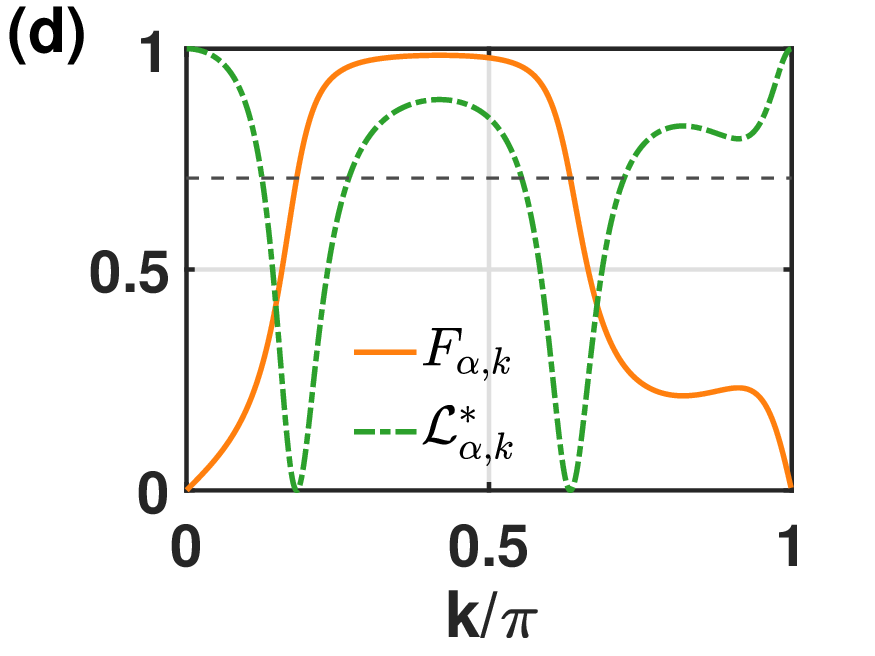}}
	{\includegraphics[width=0.24\columnwidth, height=0.18\columnwidth]{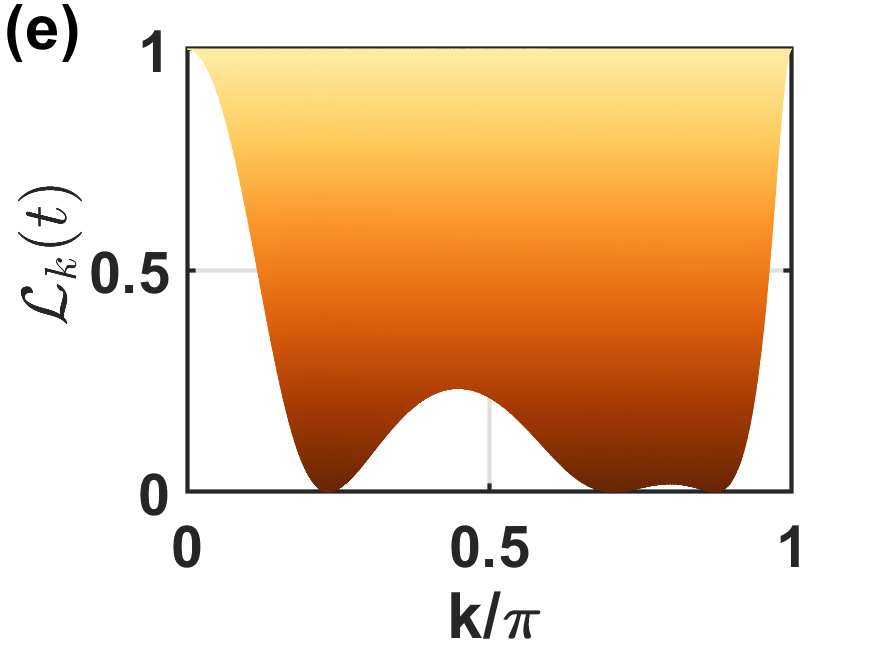}}
	{\includegraphics[width=0.24\columnwidth, height=0.18\columnwidth]{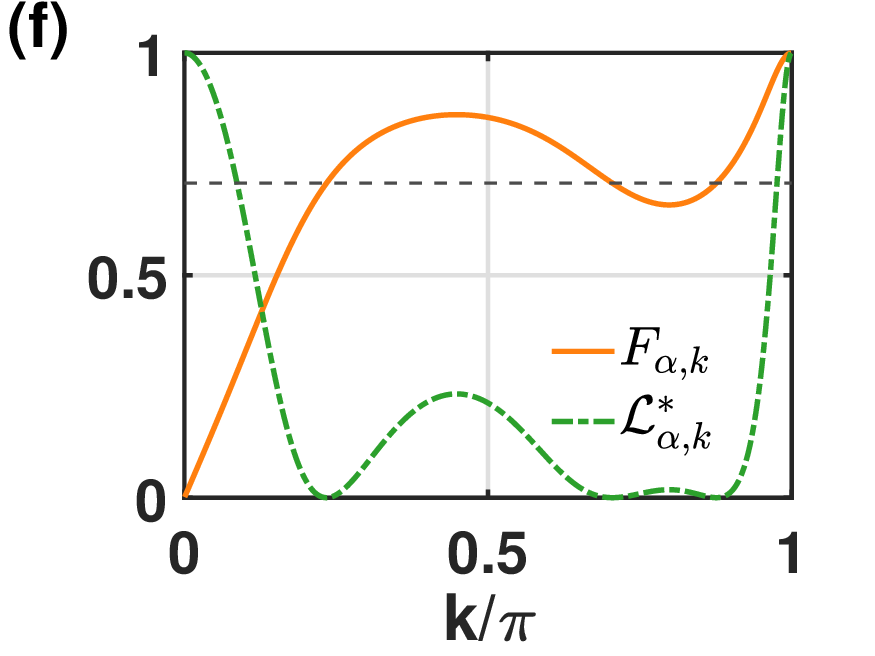}}
	{\includegraphics[width=0.24\columnwidth, height=0.18\columnwidth]{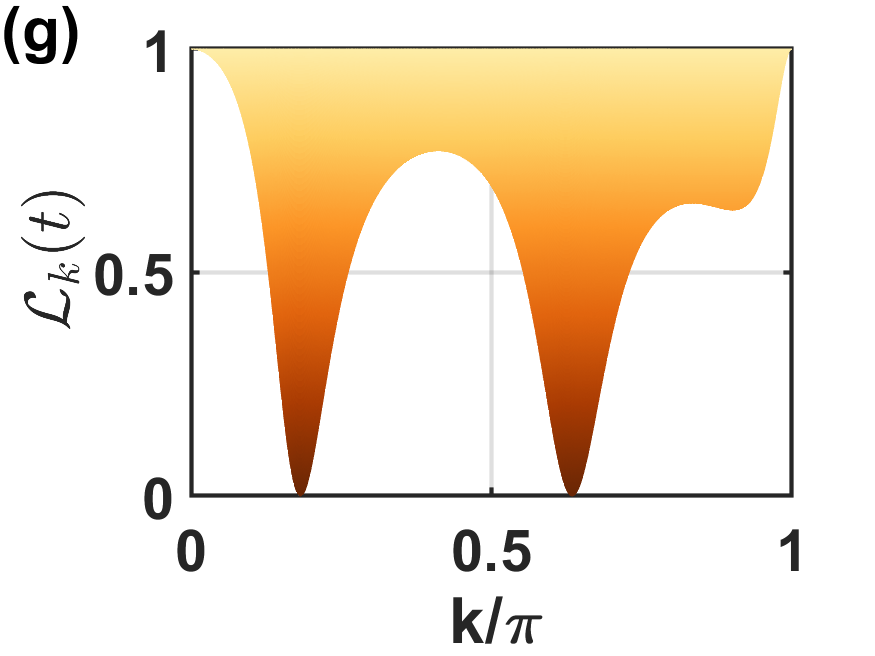}}
	{\includegraphics[width=0.24\columnwidth, height=0.18\columnwidth]{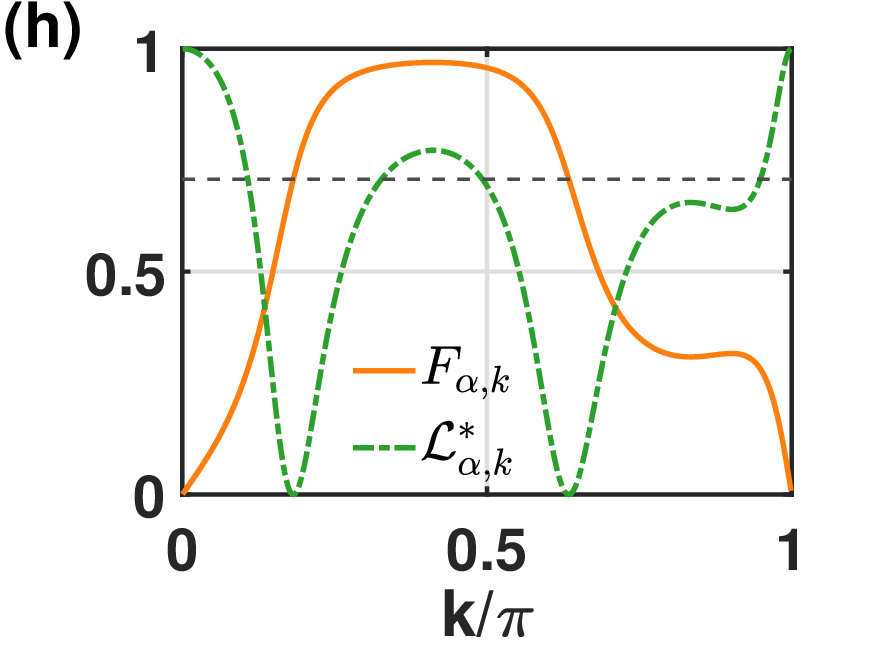}}
	\caption{ Effects of symmetric time allocation $(T_1 = T_2=\pi)$ in the periodically quenched flux process. Panels (a), (c), (e) and (g) show the side views of the time evolution of the single-mode Loschmidt echo $\mathcal{L}_k\left( t \right) $ as a function of momentum $k$, with parameters $\lambda=0.6$, $\Delta\phi=\pi/4$ for (a), $\lambda=1.6$, $\Delta\phi=\pi/2$ for (c),
	$\lambda=0.6$, $\Delta\phi=\pi$ for (e) and $\lambda=1.6$, $\Delta\phi=\pi$ for (g). Panels (b), (d), (f) and (h) display the Floquet quench fidelity $F_{\alpha,k}$ (orange solid lines) and the minimum of the single-mode Loschmidt echo $\mathcal{L}_{\alpha,k}^{*}$ (green dash–dotted lines) corresponding to (a), (c), (e) and (g), respectively, for the same driving period $T_1=T_2=\pi$. Numerical calculations were performed with $J=1$, $\gamma=1$, and $N=1000$.
	} 
	\label{fig_2}
\end{figure}

Fig.~\ref{fig_2} provides a detailed illustration of the FDQPTs for the case of equal periodic quench durations, $T_1 = T_2 = \pi$. Panels (a) and (b) show the side views of the time evolution of $\mathcal{L}_k\left( t \right) $ as a function of momentum $k$.
These panels clearly reveal that the zeros of $\mathcal{L}_k\left( t \right) $ emerge and vary with momentum, indicating the critical modes where FDQPTs occur within each driving period. 
Panels (c) and (d) display the Floquet quench fidelity $F_{\alpha,k}$ (solid orange lines) and the minimum of the single-mode Loschmidt echo $\mathcal{L}_{\alpha,k_c}^{*}$ (green dash–dotted lines) as functions of $k$ for the same driving period. 
Since $T_1 = T_2$, the two quench segments are dynamically identical, so only one set of curves is plotted for clarity, where $\alpha=1$ and $\alpha=2$ respectively denote the first and second quench segments. 
In Fig.~\ref{fig_2}(a), two critical momenta $k_c = 0.2336\pi$ and $k_c = 0.8759\pi$ are identified, where $\mathcal{L}_{\alpha,k}^{*}$ vanishes, signaling the occurrence of FDQPTs. These critical modes coincide precisely with the points where the Floquet quench fidelity $F_{\alpha,k_c} = \sqrt{2}/2$ is satisfied, 
as shown in Fig.~\ref{fig_2}(b). 
Similarly, in Fig.~\ref{fig_2}(c), two critical modes $k_c = 0.1826\pi$ and $k_c = 0.6339\pi$ can also be observed, corresponding to the momenta that fulfill the same Floquet quench fidelity condition in Fig.~\ref{fig_2}(d). Furthermore, a vertical comparison of the panels reveals that when $\lambda$ is fixed, varying $\Delta\phi$ leads to changes in both the position and number of critical momenta where FDQPTs occur. This indicates that $\Delta\phi$ plays a crucial role in inducing FDQPTs in the system.
Therefore, under symmetric time allocation ($T_1 = T_2$), critical momenta satisfying $F_{\alpha,k_c}= \sqrt{2}/2$ induce FDQPTs within the corresponding segment, where $\mathcal{L}_k\left( t \right) $ necessarily vanishes at the corresponding critical times. 

\begin{figure}
	\centering
	{\includegraphics[width=0.286\columnwidth, height=0.22\columnwidth]{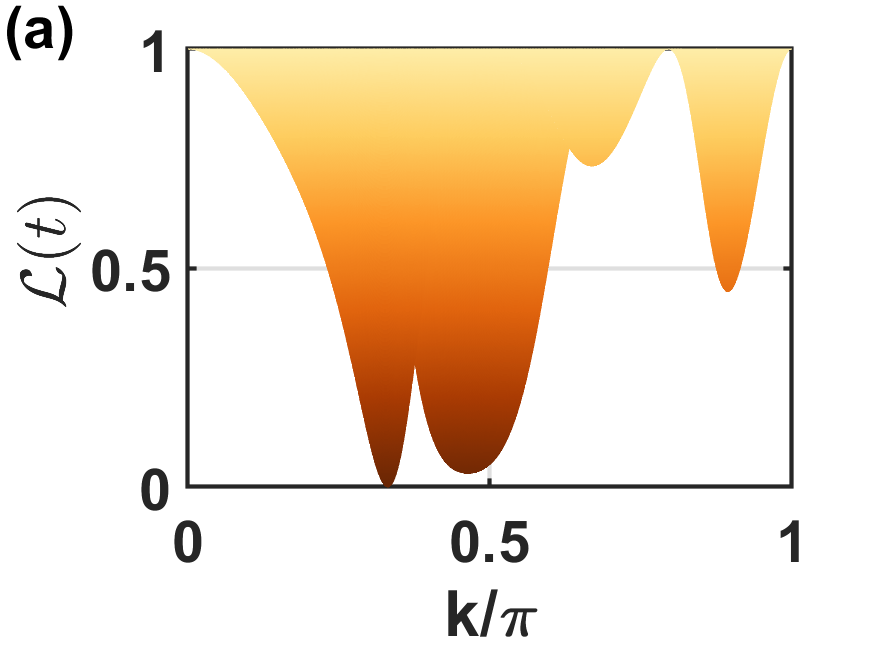}}
	{\includegraphics[width=0.286\columnwidth, height=0.22\columnwidth]{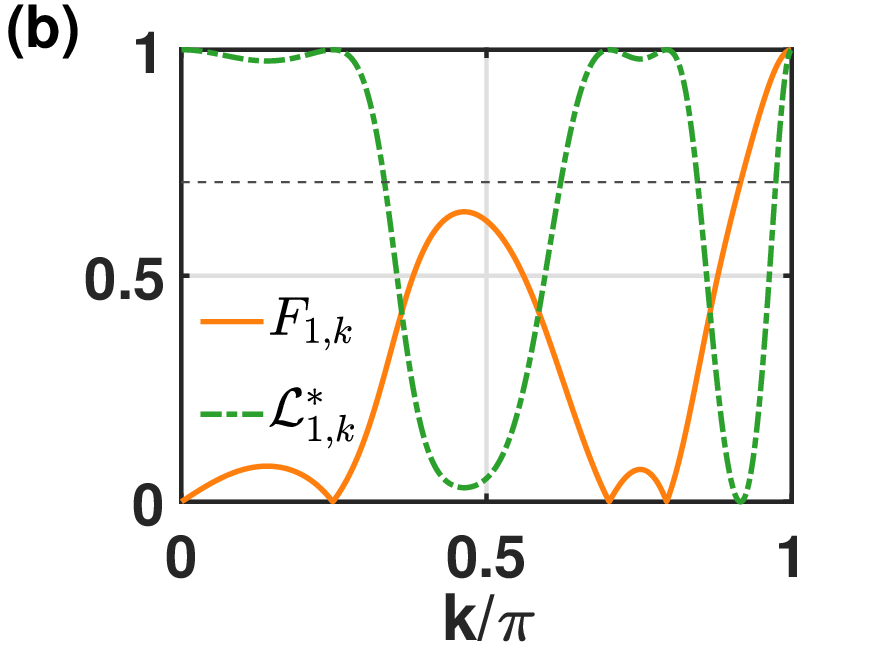}}
	{\includegraphics[width=0.286\columnwidth, height=0.22\columnwidth]{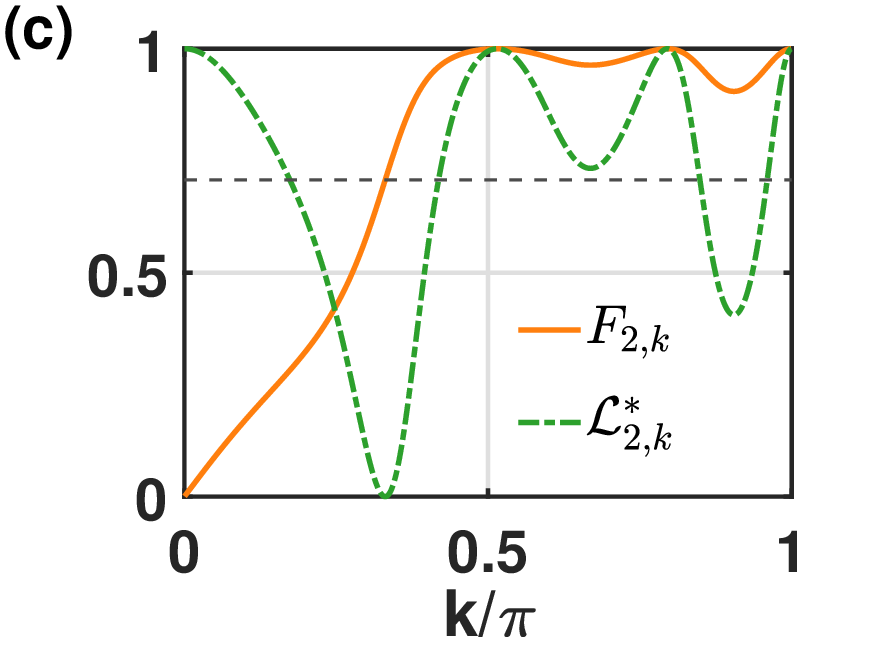}}
	{\includegraphics[width=0.286\columnwidth, height=0.22\columnwidth]{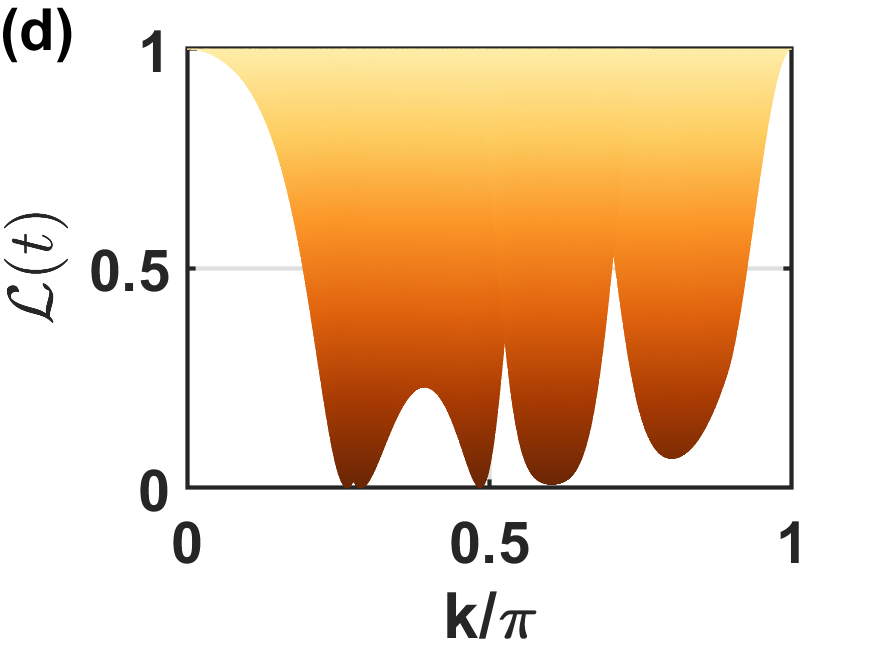}}
	{\includegraphics[width=0.286\columnwidth, height=0.22\columnwidth]{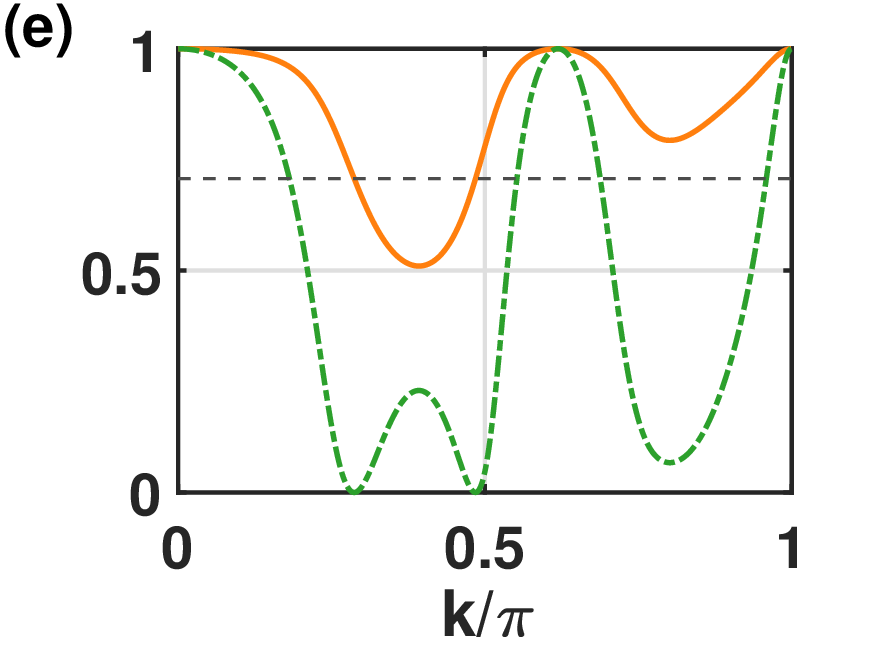}}
	{\includegraphics[width=0.286\columnwidth, height=0.22\columnwidth]{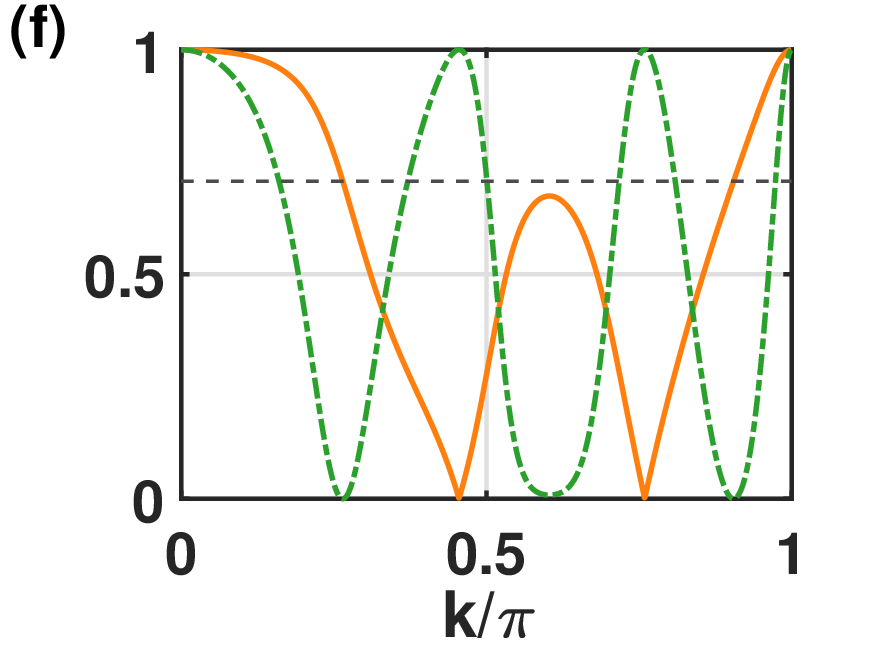}}
	\caption{Effects of asymmetric time allocation $(T_1 \ne T_2)$ in the periodically quenched flux process, where we have set $J=1$, $\gamma=1$, $N=1000$. 
	Panels (a) and (d) show the side views of the single-mode Loschmidt echo $\mathcal{L}_k\left( t \right) $ as a function of momentum $k$, with parameters $\lambda=0.8$, $\Delta\phi=\pi$, $T_1=0.8\pi$, $T_2=1.2\pi$ for (a), and $\lambda=1.2$, $\Delta\phi=\pi$, $T_1=1.2\pi$, $T_2=0.8\pi$ for (d).
	Panels (b) and (c) display the Floquet quench fidelity $F_{\alpha,k}$ (orange solid lines) and the minimum of the single-mode Loschmidt echo $\mathcal{L}_{\alpha,k}^{*}$ (green dash–dotted lines) corresponding to $t \in [0, T_1)$ and $t \in [T_1, T)$ time intervals of (a), respectively. Similarly, panels (e) and (f) show the results corresponding to the first and second time intervals of (d). 
	} 
	\label{fig_3}
\end{figure}

Fig.~\ref{fig_3} illustrates the effects of asymmetric time allocation ($T_1 \ne T_2$) on the occurrence of FDQPTs. 
In Fig.~\ref{fig_3}(a), where $T_1 = 0.8\pi$ and $T_2=1.2\pi$, the Loschmidt echo vanishes at a single critical momentum $k_c = 0.3307\pi$. 
From Fig.~\ref{fig_3}(b), one can find that the Floquet quench fidelity satisfies $F_{1,k_c}= \sqrt{2}/2$ at $k_{c,1} = 0.916\pi$, which also corresponds to $\mathcal{L}_{1,k_{c}}^{*}=0$ but does not lead to $\mathcal{L}_{1,k_{c}} = 0$.
This discrepancy arises because the corresponding critical time $t_c$, obtained from Eq. (\ref{Tc}), is not in the time interval $t \in [0, T_1)$.
Similarly, in Fig.~\ref{fig_3}(d), where $T_1=1.2\pi$ and $T_2=0.8\pi$, the Loschmidt echo reaches zero at three distinct momenta, $k_c = 0.2646\pi$, $0.2866\pi$, and $0.4847\pi$.  
By substituting these momenta into Eq.~(\ref{Tc}), one can find that the critical times corresponding to the momenta $k_{c,1} = 0.2866\pi$ and $0.4847\pi$, which satisfy $F_{1,k_c} = \sqrt{2}/2$, lie in the first segment $t \in [0, T_1)$, as shown in Fig.~\ref{fig_3}(e).
In contrast, within the second segment $t \in [T_1, T)$ (Fig.~\ref{fig_3}(f)), only $k_{c,2} = 0.2646\pi$ satisfies this condition, while another momentum $k_{c,2} = 0.905\pi$ also yields $F_{2,k_c} = \sqrt{2}/2$ but its corresponding $t_c$ falls outside this time window, and therefore no FDQPT occurs. 

Consequently, the occurrence of FDQPTs is constrained by the finite duration of each segment in the piecewise quench protocol. 
The existence of a momentum mode satisfying $F_{\alpha,k_c} = \sqrt{2}/2$ is no longer a sufficient and necessary condition for observing FDQPTs. 
To observe FDQPTs within each quench segment, both of the following requirements must be satisfied: 
(i) There exists at least one critical momentum $k_c$ such that $F_{\alpha,k_c} {=} \sqrt{2}/2$;  
(ii) The corresponding critical time $t_c$ is in the time window of that segment.
\subsection{Rate function}

The rate function can be obtained from the Loschmidt amplitude, which can also exhibit nonanalyticities within a driving period whenever FDQPTs occur, induced by the periodically flux-quenched protocol.
In the thermodynamic limit, the rate function for a two-segment quench protocol can be expressed as
\begin{equation}   
	g( t ) {=}\left\{   
	\begin{aligned}
		&{-}\int_0^{\pi}{\frac{dk}{2\pi}}\ln | \mathcal{G}_{1,k}(t)  |^2, &t \in [0, T_1), \\
		&{-}\int_0^{\pi}{\frac{dk}{2\pi}}\ln  |\mathcal{G}_{2,k} (t  )  |^2, &t \in [T_1, T). 
	\end{aligned}   
	\right. 
\end{equation}
where $\mathcal{G}_{\alpha,k}(t)$ has been defined in Eqs. (\ref{G1t}) and  (\ref{G2t}).

Fig.~\ref{fig_4} shows the time evolution of the rate function under the periodic flux-quench protocol for both the symmetric case $T_1 = T_2 = \pi$ and the asymmetric case $T_1 \ne T_2$, with all parameters chosen to match those in Fig.~\ref{fig_2} and Fig.~\ref{fig_3}. The periodic flux-quench protocol induces distinct nonanalyticities in the rate function within each driving period, and the occurrence times of these nonanalytic points coincide with the zeros of the Loschmidt echo, marking the critical times at which FDQPTs occur. 
Moreover, increasing the transverse field strength $\lambda$ results in a larger number of nonanalyticities, indicating that stronger fields induce more occurrences of FDQPTs. Obviously, the behavior of the rate function is inherited from and consistent with the features of the Loschmidt echo.
\subsection{Geometric phase and dynamical topological order parameter(DTOP)}

The features of FDQPTs in the periodic flux-quench can also appears in the Pancharatnam geometric phase (PGP), and the resulting dynamical topological order parameter (DTOP) from PGP can also consistently exhibit quantized jumps at the critical times corresponding to nonanalyticities of the rate function. \par
The geometric phase accumulated in each segment of the two-step flux-quench protocol, which is defined as
\begin{eqnarray} 
	\varPhi_{\alpha,k}^{G}(t) =\varPhi_{\alpha,k}(t)-\varPhi_{\alpha,k}^{D}(t),
\end{eqnarray} 
where $\varPhi _{\alpha,k}(t)$ denotes the total phase accumulated in segment $\alpha$, and $\varPhi _{\alpha,k}^{D}(t)$ represents the dynamical phase acquired during the system’s evolution. 
For the first time segment $t \in [0, T_1)$, the total phase is  
\begin{eqnarray}
	\varPhi _{1,k}(t)=\arctan \left( \frac{\sin \left( \xi _kt \right)  \mathbf{\hat{n}}_{\text{eff},k} \cdot \mathbf{\hat{d}}_{1,k} }{\cos \left( \xi _kt \right)} \right) 
\end{eqnarray} 
and the corresponding dynamical phase is
\begin{eqnarray}
	\varPhi _{1,k}^{D} ( t  ) &=&-\int_0^t{\left< \psi _0 ( \tau )  | H_{1,k} (\tau ) | \psi _0 ( \tau ) \right> d\tau} \nonumber \\
	&=&\xi _k ( 1-2 | u_{1,k}  |^2) t.
\end{eqnarray}
For the second segment $t \in [T_1, T)$, the total phase is 
%\begin{widetext}
	\begin{eqnarray}
		\varPhi _{2,k}=\arctan \left( \frac{\cos ( \xi _{\text{eff},k}T) \sin( \xi _k( t-T ))   \mathbf{\hat{n}}_{\text{eff},k} \cdot \mathbf{\hat{d}}_{2,k}   +\cos ( \xi _k( t-T) ) \sin ( \xi _{\text{eff},k}T )}{\cos( \xi _{\text{eff},k}T) \cos( \xi _k( t-T)) -\sin( \xi _{\text{eff},k}T) \sin  ( \xi _k ( t-T  )  )   \mathbf{\hat{n}}_{\text{eff},k} \cdot \mathbf{\hat{d}}_{2,k}} \right),   
	\end{eqnarray}
%\end{widetext}
and the corresponding dynamical phase is
\begin{eqnarray}
	\varPhi _{2,k}^{D} (t  ) &=&-\int_0^t{\left< \psi _0 ( \tau  ) \,|\,H_{2,k} (\tau  ) |\psi _0 ( \tau  ) \right>  d\tau} \nonumber \\
	&=&\xi _k ( 1-2 | u_{2,k} |^2  ) t. 
\end{eqnarray}
Combining the contributions from both segments, the DTOP under the periodic flux-quench protocol is expressed as
\begin{eqnarray}
	\nu(t)=\frac{1}{2\pi}\int_0^{\pi}{\frac{\partial \varPhi _{\alpha,k}^{G}\left(t \right)}{\partial k}}dk.
\end{eqnarray}
The quantized jumps of the DTOP in each segment of the piecewise quench evolution correspond to the occurrence of FDQPTs.
\begin{figure}
	\centering
	{\includegraphics[width=0.24\columnwidth, height=0.18\columnwidth]{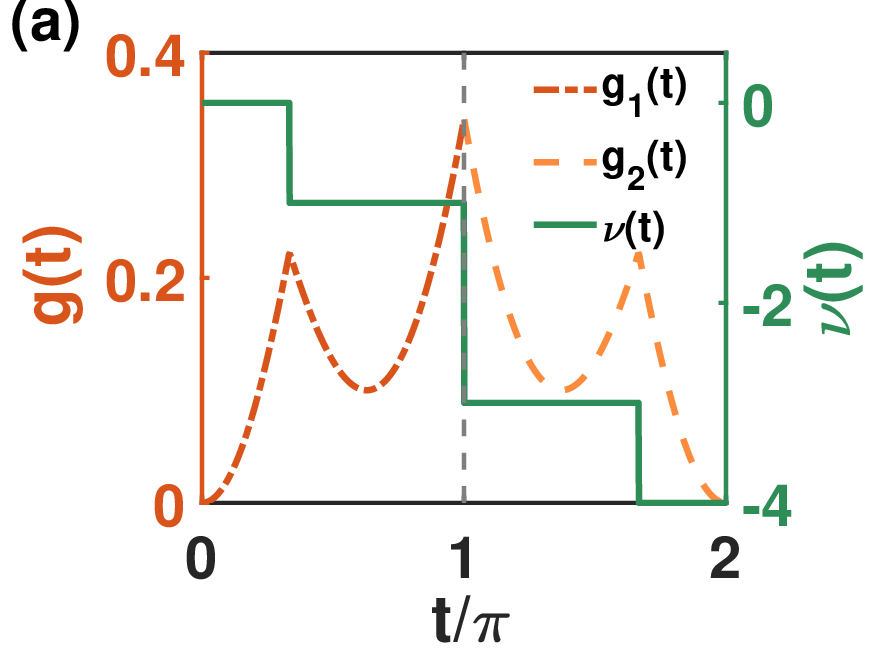}}
	{\includegraphics[width=0.24\columnwidth, height=0.18\columnwidth]{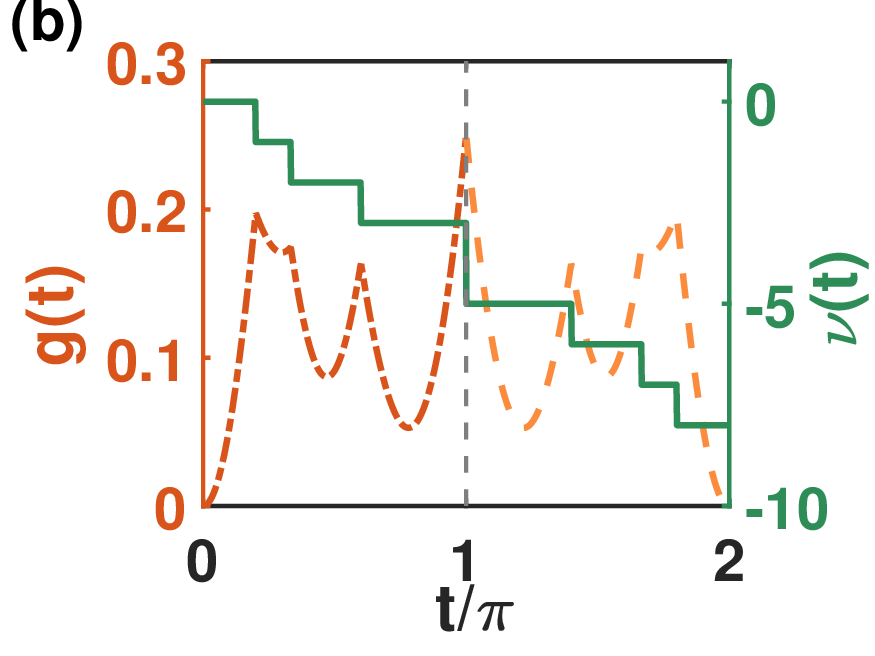}}
	{\includegraphics[width=0.24\columnwidth, height=0.18\columnwidth]{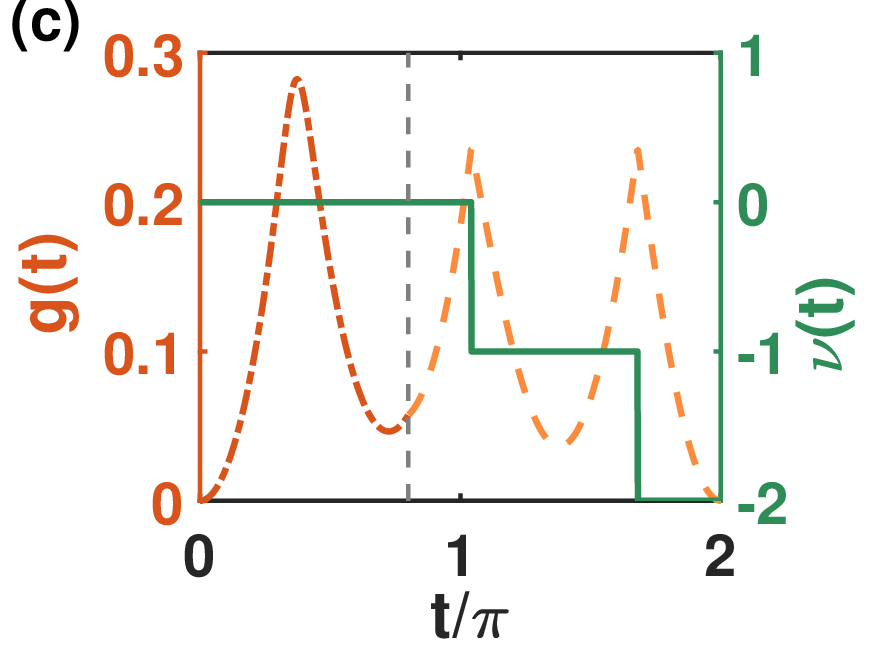}}
	{\includegraphics[width=0.24\columnwidth, height=0.18\columnwidth]{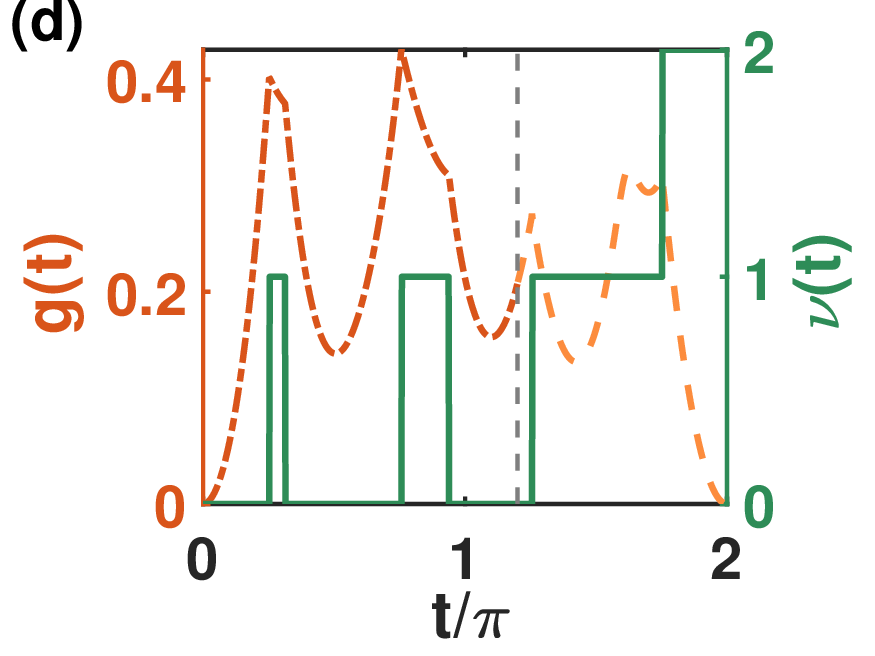}}
	\caption{Rate function $g(t)$ and DTOP $\nu(t)$ for the two-segment quench protocol. The red dash–dotted and orange dashed curves represent the rate functions $g_1(t)$ and $g_2(t)$, corresponding to the first and second segments of the flux-quench, respectively. The green solid curve denotes the $\nu(t)$. The grey vertical dashed line marks the boundary between the two quench segments. Numerical calculations were performed with $J=1$, $\gamma=1$, and $N=1000$. 
		(a) $\lambda=0.6$, $\Delta\phi=\pi/4$, $T_1=T_2=\pi$;
		(b) $\lambda=1.6$, $\Delta\phi=\pi/2$, $T_1=T_2=\pi$;
		(c) $\lambda=0.8$, $\Delta\phi=\pi$, $T_1=0.8\pi$, $T_2=1.2\pi$;
		(d) $\lambda=1.2$, $\Delta\phi=\pi$, $T_1=1.2\pi$, $T_2=0.8\pi$.
	} 
	\label{fig_4}
\end{figure}

Fig.~\ref{fig_4} clearly shows that the DTOP (green solid line) exhibits quantized jumps under different choices of system parameters. The times at which these jumps occur match precisely the locations of the nonanalytic points in the rate function discussed in the previous section, thereby indicating that the system undergoes FDQPTs at those moments.
In particular, as shown in Fig.~\ref{fig_4} (a) and (b), where the two quench segments satisfy $T_1=T_2$, the Hamiltonian switches from $H_1$ to $H_2$ and induces FDQPTs, which is consistent with the integer changes of DTOP. This occurs because the critical time is $t_c=T$, at which two distinct critical momenta $k_c$ appear.
In addition, as shown in Fig.~\ref{fig_4} (c) and (d), the cusp-like features of the rate function may not always indicate the emergence of FDQPT, whereas the integer-quantized jumps of the DTOP provide a reliable diagnostic signature, i.e., the system undergoes an FDQPT only when the DTOP exhibits an integer-quantized change. 

\subsection{FDQPTs expressed on the Bloch sphere}

Since the time evolution of the periodically quenched system can be mapped onto a trajectory on the surface of the Bloch sphere, where the emergence of FDQPTs may be reflected by the geometric properties of these evolution trajectories, the temporal constraint for the occurrence of FDQPTs under periodic driving can also be captured by the evolution trajectories on the Bloch sphere.

Under the micromotion dynamics induced by periodic driving, the density matrix for each $k$-mode is initially given by 
\begin{eqnarray}
	\rho_k \left(0 \right) =\rho _{\text{eff},k} =\frac{1}{2}\left[ \,1-\mathbf{\hat{n}}_{\text{eff},k}  \cdot \boldsymbol{\sigma} \right] . 
\end{eqnarray}
At an arbitrary time $t$, the density matrix can be generally expressed as
\begin{eqnarray}
	\rho_k ( t ) =\frac{1}{2} [1-\mathbf{\hat{d}}_k(t) \cdot \boldsymbol{\sigma } ],
\end{eqnarray}
where we have defined $\mathbf{\hat{d}}_k  =\mathbf{ {d}}_{k}/|\mathbf{ {d}}_{k}|$.

During the first stage $t \in [0, T_1)$ of the driving protocol, the Bloch vector evolves as
\begin{eqnarray}
	&&\mathbf{\hat{d}}_k ( t  )=\mathbf{\hat{n}}_{\text{eff},k}\cos  ( 2|\mathbf{d}_{1,k}|t  ) {+}2\mathbf{\hat{d}}_{1,k} ( \mathbf{\hat{n}}_{\text{eff},k}\cdot \mathbf{\hat{d}}_{1,k} ) \sin ^2 ( |\mathbf{d}_{1,k}|t  )  -\mathbf{\hat{n}}_{\text{eff},k}\times \mathbf{\hat{d}}_{1,k}\sin  ( 2|\mathbf{d}_{1,k}|t  ),
\end{eqnarray}
and in the second stage $t \in [T_1, T)$, the evolution is governed by
\begin{eqnarray}
	&&\mathbf{\hat{d}}_k(t) =\mathbf{\hat{d}}_{\text{T}_1,k}\cos  ( 2|\mathbf{d}_{2,k}|t  ) {+}2\mathbf{\hat{d}}_{2,k} ( \mathbf{\hat{d}}_{\text{T}_1,k}\cdot \mathbf{\hat{d}}_{2,k}  ) \sin ^2 ( |\mathbf{d}_{2,k}|t  ) -\mathbf{\hat{d}}_{\text{T}_1,k}\times \mathbf{\hat{d}}_{2,k}\sin ( 2|\mathbf{d}_{2,k}|t), 
\end{eqnarray}
where %$\mathbf{\hat{d}}_{\text{T}_1}\equiv \mathbf{\hat{d}}\left( k,T_1 \right)$ 
$\mathbf{\hat{d}}_{\text{T}_1,k}$ represents the Bloch vector at the end of the first micromotion interval and serves as the initial condition for the second stage of the evolution.
\begin{figure}
	\centering
	{\includegraphics[width=0.33\columnwidth, height=0.25\columnwidth]{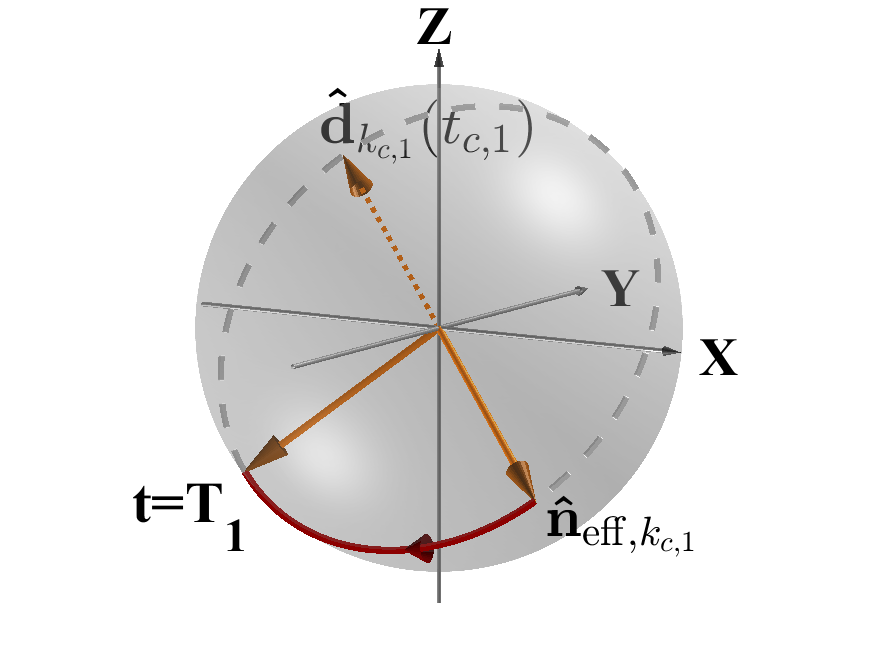}}
	{\includegraphics[width=0.33\columnwidth, height=0.25\columnwidth]{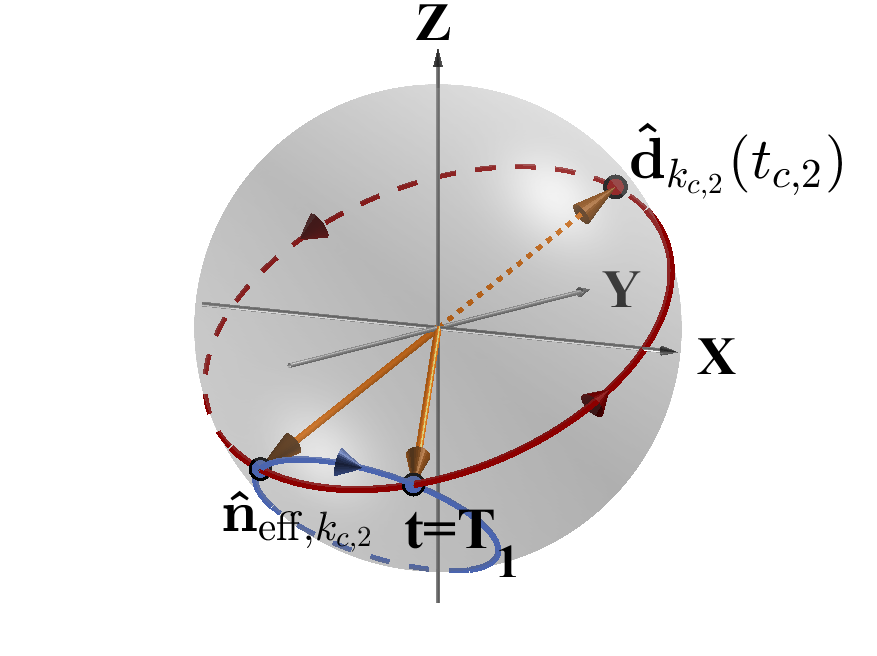}}
	\caption{Schematic illustration of the micromotion dynamics of the Bloch vector $\mathbf{\hat{d}}_k(t) $ on the Bloch sphere within one driving period $T$, where we have set  $\lambda=0.8$, $\Delta\phi=\pi$ , $T_1=0.8\pi$, $T_2=1.2\pi$ in the calculation. 
		(a) At the critical momentum $k_{c,1}$, the evolution trajectory of $\mathbf{\hat{d}}_k(t)$ starting from the initial $\mathbf{\hat{n}}_{\text{eff},k}$ precesses around the axis $\mathbf{\hat{d}}_{1,k}$ for $t \in [0, T_1)$ (red solid line). The gray dashed line represents the incomplete trajectory that cannot be realized for $t>T_1$ due to the finite finite duration of the driving stage. 
		(b) At the critical momentum $k_{c,2}$, $\mathbf{\hat{d}}_k(t)$ precesses around $\mathbf{\hat{d}}_{1,k}$ for $t \in [0, T_1)$ (blue line) and subsequently around $\mathbf{\hat{d}}_{2,k}$ for $t \in [T_1, T)$ (red line). The blue circle marks the state at $t=T_1$, and the red circle denotes the critical point at which $\mathbf{\hat{d}}_{k_c}\left(t_c \right)$ becomes antiparallel to the initial Bloch vector $\mathbf{\hat{n}}_{\text{eff},k}$, signaling the occurrence of FDQPTs.
	} 
	\label{fig_5}
\end{figure}

For a system after a single quench, the DQPT occurs when the evolved state become orthogonal to the initial state in a single quench, i.e., the Bloch vector satisfies $\mathbf{\hat{d}}_{k_c}(t_c ) =-\mathbf{\hat{d}}_{i,k_c}$
\cite{guo2019observation,PhysRevB.97.060304}. Since the system can evolve long enough to reach the critical time $t_c$, where the Bloch vector $\mathbf{\hat{d}}_{k_c} $ becomes antiparallel to the initial direction, a DQPT is guaranteed to occur. Thus, the critical momentum condition is both necessary and sufficient for the phenomenon to occur. However, for the periodically driven system, the evolution within each segment is constrained by both $T_1$ and $T_2$. As shown in Fig.~\ref{fig_5}(a), even if a critical momentum $k_{c,1}$ satisfies $F_{1,k_c} = \sqrt{2}/2$, the corresponding critical time may fall outside the time window $t \in [0, T_1)$ of that segment. In this case, the Bloch vector does not reach the antiparallel direction before the Bloch vector $\mathbf{\hat{d}}_k\left(t \right)$ rotates around a new axis $\mathbf{\hat{d}}_{2,k}$ and starts a new evolution trajectory. In Fig.~\ref{fig_5}(b), the red trajectory shows that within the time interval $t \in [T_1, T)$, the Bloch vector $\mathbf{\hat{d}}_k\left(t \right)$ reaches the antiparallel position relative to the initial Bloch vector $\mathbf{\hat{n}}_{\text{eff},k} $ at two distinct moments, indicating the occurrence of FDQPTs. Therefore, the emergence of FDQPTs within each driving segment requires that the critical momentum satisfies the Floquet fidelity condition, and the associated critical time $t_c$ must fall within the evolution time of the corresponding segment. In addition, by using $ ( \boldsymbol{\sigma }\cdot \mathbf{\hat{n}}_{\text{eff},k}  )  ( \boldsymbol{\sigma }\cdot \mathbf{\hat{d}}_{\alpha,k}  ) =\mathbf{\hat{n}}_{\text{eff},k}\cdot \mathbf{\hat{d}}_{\alpha,k}\,\mathbb{I}+i\boldsymbol{\sigma }\cdot  ( \mathbf{\hat{n}}_{\text{eff},k}\times \mathbf{\hat{d}}_{\alpha,k}  ) $, we can find that the occurrence of FDQPTs requires $\left\{ H_{\text{eff},k_c}, H_{\alpha,k_c} \right\} =0$ for the effective Floquet Hamiltonian and the segmental quench Hamiltonians, which provides a new perspective for exploring FDQPTs.

\section{conclusion}\label{sec5} 

We have investigated FDQPTs in an extended XY spin chain subjected to periodic flux quench, showing that the emergence of FDQPTs depends on the flux difference $\Delta\phi=\phi_1-\phi _2$ between the two quench segments, while the increase in transverse field strength will give rise to multiple FDQPTs within a single driving period.
By introducing the concept of Floquet quench fidelity under periodic driving, we discuss the necessary and sufficient conditions for the occurrence of FDQPTs with both symmetric ($T_1=T_2$) and asymmetric ($T_1\neq T_2$) time allocations.
Unlike the case of single-quench, the periodically driven piecewise quench protocol imposes an additional temporal constraint, i.e., the FDQPT 
can be observed only if the critical time associated with a momentum mode $k_c$, which satisfies $F_{\alpha,k_c}{=} \sqrt{2}/2$, is in the evolution time window of the corresponding segment, ensuring that the Bloch vector reaches the antiparallel direction before $\mathbf{\hat{d}}_k(t) $ rotates around a new axis $\mathbf{\hat{d}}_{\alpha,k}$ and starts a new evolution trajectory.
Such requirements for the emergence of FDQPTs are also valid for other parameter-controlled periodic quenched systems.
Our results clarify how periodic flux-quench protocols affect nonequilibrium critical behavior in driven quantum many-body systems, offering a foundation upon which more general classes of periodically driven models may be analyzed and compared.

\section*{CRediT authorship contribution statement}
Wen-Hui Nie: Writing–original draft, Software, Methodology, Formal analysis, Conceptualization, Visualization. Mei-Yu Zhang: Validation,
Investigation. Lin-Cheng Wang: Writing–review, editing, Validation, Conceptualization. Chong Li: Conceptualization, Investigation.

\section*{Declaration of competing interest}
This manuscript has not been published before and is not being considered for publication elsewhere. All authors have contributed to the creation of this manuscript for important intellectual content and read and approved the final manuscript. We declare there is no conflict of interest.

\section*{Acknowledgement}
We would like to thank Zheng Liu for helpful discussions. This work was supported by National Natural Science Foundation of China (NSFC) under grant No. 11475037, No. 11574041.

\section*{Data availability}
Data will be made available on request.

%\newpage
\footnotesize
\setlength{\bibsep}{0pt plus 0.3ex}
\bibliography{references}

\end{document}